\documentclass[aps,superscriptaddress,prb,amsmath,amssymb,reprint,longbibliography]{revtex4-2}

\usepackage[USenglish]{babel}
\usepackage[T1]{fontenc}
\usepackage[utf8]{inputenc}
\usepackage{graphicx}
\usepackage{upgreek}
\usepackage[unicode=true,colorlinks=true,citecolor=blue,urlcolor=blue]{hyperref}
\usepackage{multirow}
\usepackage{bm}
\usepackage[normalem]{ulem}

\usepackage{eufrak}
\usepackage{color}
\usepackage{physics}

\newcommand{\nix}[1]{}
\let\oldsec\section

\newcommand{\e}{\mathrm{e}}
\renewcommand{\k}{\textbf{k}}
\renewcommand{\i}{{\rm i}}
\renewcommand{\d}{\mathrm d}
\renewcommand{\expval}[1]{\left\langle #1 \right\rangle}

\newcommand{\addDima}[1]{#1}
\newcommand{\commentDima}[1]{}

\begin{document}

\title{Tuning the nuclei-induced spin relaxation of localized electrons\\ by the quantum Zeno and anti-Zeno effects}

\date{\today}
	
\author{V.~Nedelea}
\affiliation{Experimentelle Physik 2, Technische Universit\"at Dortmund, 44221 Dortmund, Germany}
	
\author{N.~V.~Leppenen}
\affiliation{Ioffe Institute, Russian Academy of Sciences, 194021 St.\,Petersburg, Russia}

\author{E.~Evers}
\affiliation{Experimentelle Physik 2, Technische Universit\"at Dortmund, 44221 Dortmund, Germany}

\author{D.~S.~Smirnov}
\email{smirnov@mail.ioffe.ru}
\affiliation{Ioffe Institute, Russian Academy of Sciences, 194021 St.\,Petersburg, Russia}

\author{M.~Bayer}
\affiliation{Experimentelle Physik 2, Technische Universit\"at Dortmund, 44221 Dortmund, Germany}

\author{A.~Greilich}
\email{alex.greilich@tu-dortmund.de}
\affiliation{Experimentelle Physik 2, Technische Universit\"at Dortmund, 44221 Dortmund, Germany}

\begin{abstract}  
Quantum measurement back action is fundamentally unavoidable when manipulating electron spins. Here we demonstrate that this back action can be efficiently exploited to tune the spin relaxation of localized electrons induced by the hyperfine interaction. In optical pump-probe experiments, powerful probe pulses suppress the spin relaxation of electrons on Si donors in an InGaAs epilayer due to the quantum Zeno effect. By contrast, an increase of the probe power leads to a \addDima{speed up of the spin relaxation} for electrons in InGaAs quantum dots due to the quantum anti-Zeno effect. The microscopic description shows that the transition between the two regimes occurs when the spin dephasing time is comparable to the probe pulse repetition period.
\end{abstract}

\maketitle

Modification of quantum system dynamics due to the interaction with a measurement apparatus can always be described microscopically, for example, based on the Schr\"odinger equation~\cite{Frerichs1991,Facchi2001}. In many cases, however, the general concepts of strong or weak measurements~\cite{Kraus1983,Braginsky1996,Haroche2006,Clerk2010} can be applied. The former, also known as von Neumann type of measurements, is widely discussed nowadays for measurement-based quantum computation~\cite{Raussendorf2003,Briegel2009,Benjamin2009}, while weak measurements are often implemented experimentally to minimize the system perturbation.

\addDima{Frequent measurements} can lead to freezing of the quantum dynamics, known as quantum Zeno effect~\cite{Khalfin1958,Misra1977}, which requires measurements with a repetition period $T_R$ shorter than the Zeno time~\cite{facchi_2008}, $\tau_Z$, the time of non-Markovian relaxation. The less known and less \addDima{universal is} the quantum anti-Zeno effect which is the acceleration of the system relaxation due to the quantum back action~\cite{Kaulakys1997,Facchi2001a,Maniscalco2006}. It can occur when the repetition period and the Zeno time are of the same order, $T_R\sim\tau_Z$~\cite{facchi_2008}. \addDima{In fact, this condition can be easily realized}, but often the measurement involves additional heating and other perturbations, from which the quantum anti-Zeno effect can be challenging to separate.

The quantum Zeno effect is important for quantum information processing~\cite{FransonQZEGate2004,HuangQZEGate2008}, especially with spin-based qubits, as it can be used to increase the electron spin relaxation time, the quantum anti-Zeno effect, by contrast, allows one to quickly erase spin polarization so that both effects should be taken into account when measuring spin qubits. However, the short Zeno time of free charge carriers strongly hinders reaching the regime of the quantum Zeno effect.

For localized electrons, the spin relaxation mechanisms~\cite{Dyakonov2017} related to orbital electron motion, such as Elliot-Yafet~\cite{Elliott1954,Yafet1963}, Bir-Aronov-Pikus~\cite{Bir1975}, and Dyakonov-Perel~\cite{Dyakonov1972}, are suppressed. \addDima{Then,} the hyperfine interaction with the lattice nuclei plays the dominant role in electron spin relaxation~\cite{PRBMerkulov02,Khaetskii2002}. Due to the long nuclear spin coherence times, the electron spin relaxation is hyperfine-induced and is essentially non-Markovian~\cite{Coish2004,Coish2009,Shumilin2021}. In this case, the Zeno time equals the electron spin dephasing time $T_2^*$ making the quantum Zeno effects particularly important for localized electron spins~\cite{Poltavtsev2014,ZenoPRB,Leppenen2022}. As quantum dots represent attractive candidates for scalable quantum information processing, manifestations of the quantum Zeno effects were reported: suppression of tunneling~\cite{Hackenbroich1998,Khomitsky2012,Kang2017,AhmadiniazPRR}, stabilization of optical emission~\cite{Yamaguchi:08,XuQZEGate2009,NutzPRA2019}, and nuclear spin freezing~\cite{Klauser2008,mkadzik2020controllable,PhysRevLett.126.120603}. None of the previous studies, surprisingly, has addressed Zeno effects for the nuclei-induced spin relaxation of localized carriers experimentally.

In our work, we apply the optical pump-probe technique to manipulate the spin relaxation time of electrons localized in quantum dots or on donors, utilizing the quantum Zeno and anti-Zeno effects. Due to the scaling of the dephasing time with the localization volume, $T_2^*\propto\sqrt{V}$, we observe the opposite Zeno effects for these two systems for similar experimental conditions.

\textit{Experiment.} 
The first sample consists of 20 layers of $n$-doped InGaAs self-assembled quantum dots (QDs) separated by $60$-nm barriers of GaAs. A $\delta$-doping of Si 16\,nm above each layer provides a single electron per QD on average. The QD density per layer is about $10^{10}\,$cm$^{-2}$. Rapid thermal annealing at 945\,$^{\circ}\text{C}$ for 30\,s shifts the average emission energy to 1.39\,eV and reduces the spread of the QD size distribution~\cite{PRBSchering21}. The second sample consists of a $10\,\mu$m thick InGaAs epilayer with an indium fraction of 3\,\%. It was grown by molecular beam epitaxy on a GaAs substrate. The sample was doped with Si atoms providing a donor carrier density of $3.9\times 10^{16}\,$cm$^{-3}$~\cite{PRBClara22}. 

We use the pump-probe technique to measure the spin dynamics using the Faraday rotation technique. We apply a longitudinal magnetic field parallel to the optical $z$ axis (Faraday geometry). Localized spin-singlet trion complexes are resonantly excited for the QD sample and on the low-energy PL flank for the epilayer sample. The laser pulses with a duration of 2\,ps are emitted with $T_R=1$\,GHz repetition frequency. Samples are cooled to a temperature of 6\,K in a cryostat. The beams were focused on the sample into a spot with 50\,$\mu$m diameter for the pump and 45\,$\mu$m diameter for the probe, for which we measured the Faraday ellipticity. See supplemental material section S1.A for further technical details and optical properties of samples.

\begin{figure}[t]
  \includegraphics[width=\columnwidth]{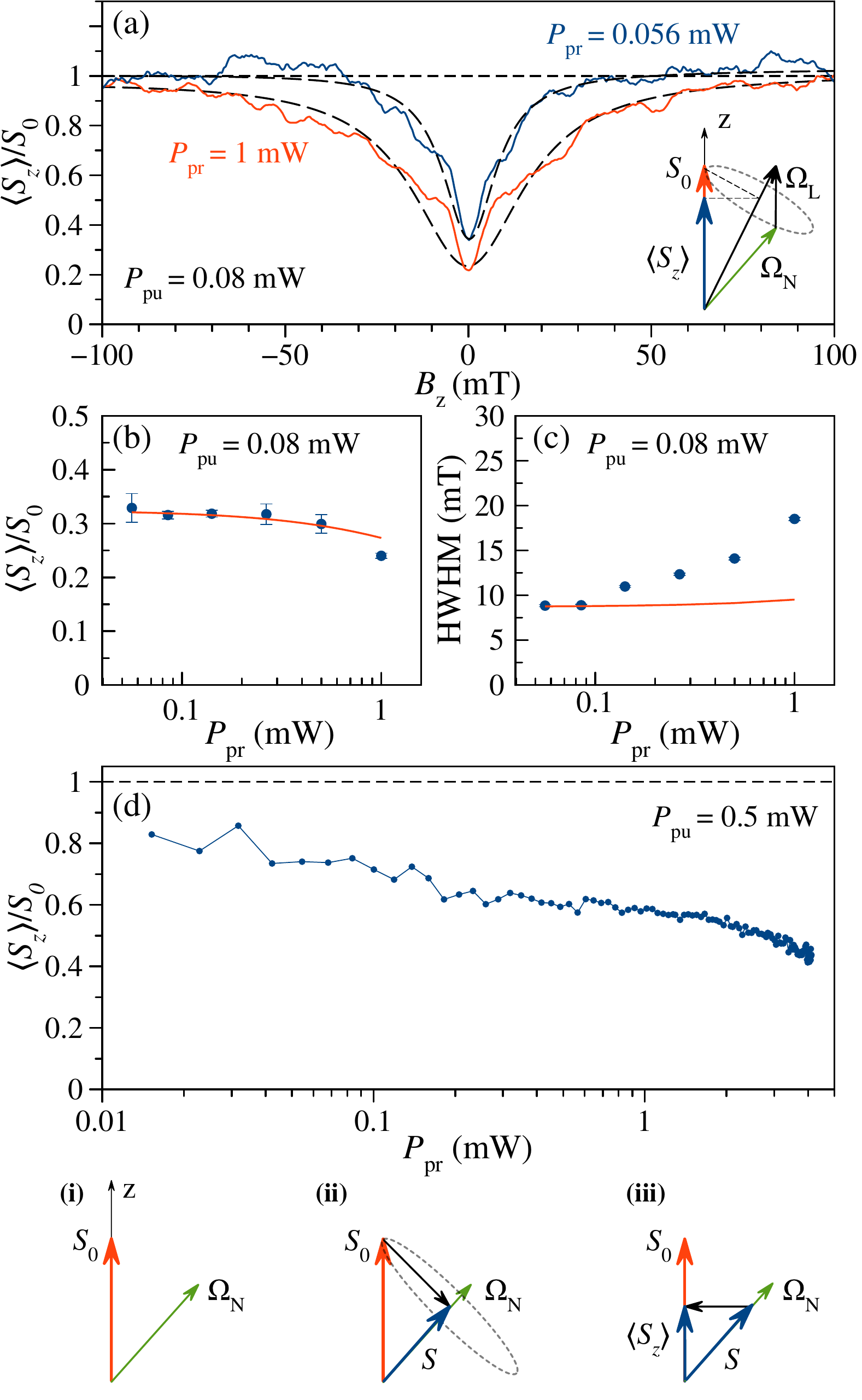}
  \caption{(a) Two exemplary PRCs for the QD sample with the pump and probe powers given by the labels and Lorentzian fits shown by the black dashed curves. The inset shows the mechanism of polarization recovery in Faraday geometry. Blue dots in panels (b) and (c) show the extracted values of the normalized zero field spin polarization and the HWHM of the PRCs, respectively. Orange lines are the theory fit curves. (d) Same as (b), but for $P_\text{pu}=0.5\,$mW. The sketches below illustrate the microscopic origin of the anti-Zeno effect, see the theory part for an explanation. $T=6\,$K. \commentDima{In (b) it would be nice to change the vertical scale to make the power dependence more clear.}}
  \label{fig:1}
\end{figure}

\textit{Results.} Figure~\ref{fig:1}(a) shows the spin polarization $\langle S_z \rangle$ of the QD sample as a function of the magnetic field for $-50\,$ps delay between pump and probe pulses, identical to $+950 \,$ps delay. The pump pulse is weak with $P_\text{pu}=0.08\,$mW average power, using two probe power settings which define the measurement strength: $P_\text{pr}=0.056\,$mW (blue curve) and $1\,$mW (\addDima{orange} curve). Each trace normalized to its value $S_0$ in 300\,mT high magnetic field. At zero magnetic field, the spin polarization is reduced by the hyperfine interaction with the fluctuating nuclear spin bath of the host lattice~\cite{Glazov2018}. The longitudinal magnetic field $B_z$ decouples the electron spins from the nuclear environment and increases the spin polarization, leading to the polarization recovery curves (PRC) shown in Fig.~\ref{fig:1}(a). The inset in Fig.~\ref{fig:1}(a) shows a sketch of the spin stabilization: the projection of the average spin polarization $\langle S_z \rangle$ on the $z$-axis rises to $S_0$ with increasing longitudinal magnetic field, shown by $\Omega_\text{L}$. $\Omega_\text{N}$ marks the average nuclei-induced Overhauser field.

The PRCs can be approximately fit by Lorentzians, see the black dashed lines. From the fits, we determine the corresponding amplitudes and half widths at half maximum (HWHM), which we show in Figs.~\ref{fig:1}(b) and~\ref{fig:1}(c), respectively, as function of the probe power. For small probe power, the HWHM of $8$\,mT with the electron longitudinal $g$-factor $g_e=-0.64$~\cite{PRBSchering21} reveals an electron spin dephasing time $T_2^*=3.5$\,ns of the order of the pulse repetition period $T_R=1$\,ns. The relative PRC depth in this limit is 2/3, as expected for isotropic nuclear spin fluctuations suppressing the spin polarization with $2/3$ probability if they are in the $(xy)$ plane and not affecting the spin polarization with $1/3$ probability if they are along the $z$ axis~\cite{PRBMerkulov02}. With increasing measurement strength (probe power), the HWHM and depth of the PRC increase. The latter reveals the shortening of the nuclei-induced spin relaxation time due to the quantum anti-Zeno effect.

In Fig.~\ref{fig:1}(d), we demonstrate measurements for the higher pump power of 0.5\,mW in a wider range of probe powers, supporting the previous observations. In this case, we scan the probe power at two fixed magnetic fields, $B_z=0$ for $\expval{S_z}$ and $B_z=300$\,mT for $S_0$, and plot the ratio of the detected spin polarizations~\footnote{The vertical offset of the values in Fig.~\ref{fig:1}(d) relative to panel (b) is related to the different pump power. The higher pump power in these measurements is a compromise between the noise level at very low pump and probe powers and the effect of the pump on the spin polarization.}.

\begin{figure}
\includegraphics[width=\columnwidth]{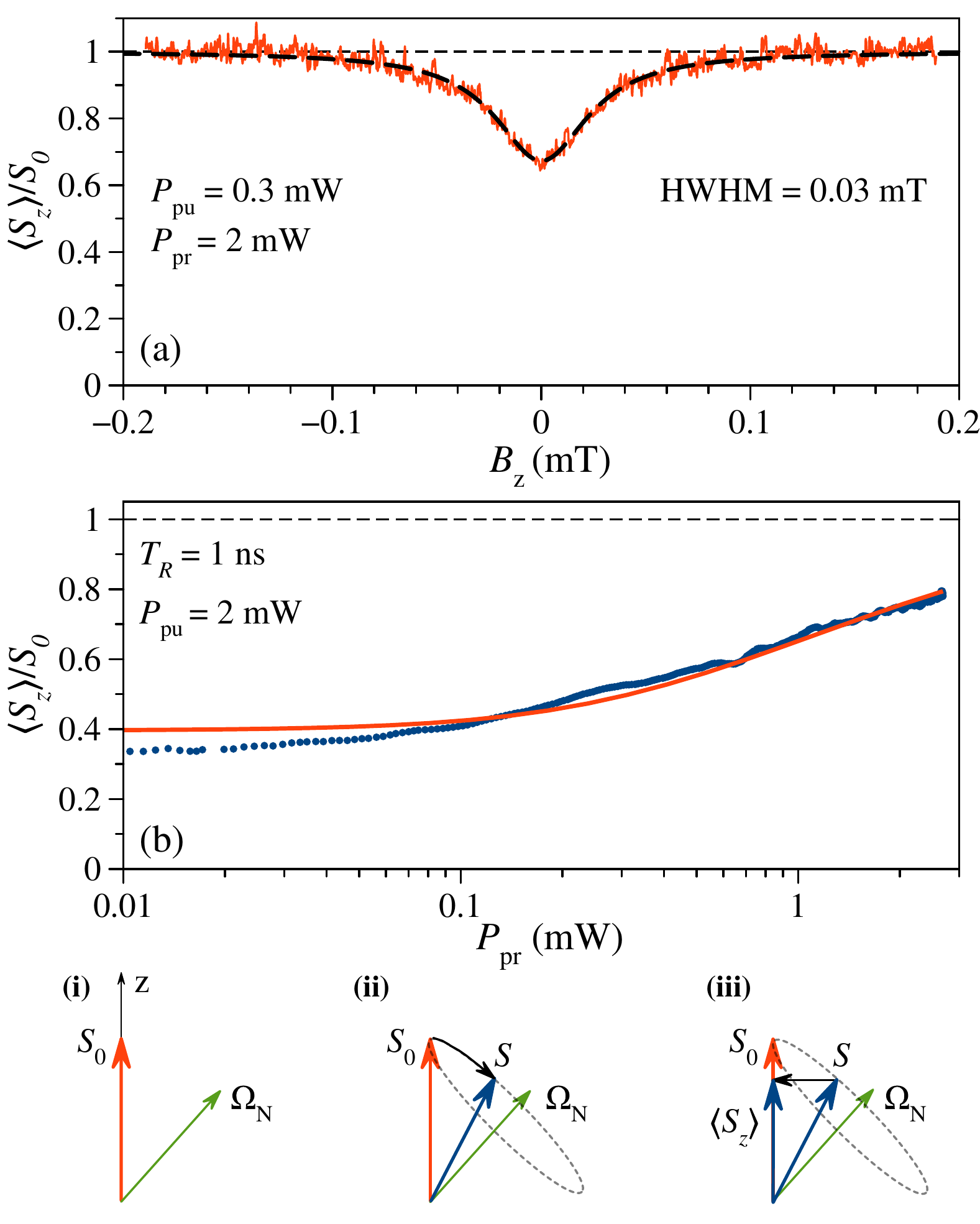}
\caption{(a) Exemplary PRC for the epilayer sample with the pump and probe powers stated by the labels. The black dashed curve gives the Lorentzian fit with the HWHM of 0.03\,mT. (b) Probe power dependence of the normalized spin polarization (blue dots) with its theory fit (orange curve). The sketches illustrate the microscopic mechanism of the quantum Zeno effect, see the theory part for an explanation.}
\label{fig:2}
\end{figure}

In the epilayer sample, we observe the opposite behavior using the same method. Figure~\ref{fig:2}(a) shows a PRC example for $P_{\text{pu}}=0.3$\,mW and $P_{\text{pr}}=2$\,mW, together with a Lorentzian fit. The HWHM of the PRC, in this case, is only $0.03$\,mT, which is smaller than for the QD sample by more than two orders of magnitude, c.f. Fig.~\ref{fig:1}(a). This is related to the larger localization volume of electrons on shallow donors than in QDs, the resulting weaker role of the hyperfine interaction, and, accordingly, the longer electron spin dephasing time $T_2^*$~\cite{PRBClara22}. Further, Fig.~\ref{fig:2}(b) demonstrates the probe power dependence of the spin polarization $\expval{S_z}$ at zero magnetic fields normalized to $S_0$ at the saturation value of 3\,mT. In stark contrast to Fig.~\ref{fig:1}(d), the normalized spin polarization rises with increasing measurement strength, revealing the quantum Zeno effect.

The observation of opposite effects for the measurement impact in the two samples for the same probe pulse repetition period and similar other conditions calls for a theoretical foundation.

\textit{Theory.} The spin dynamics of localized electrons \addDima{between the pulses} in the presence of hyperfine interaction with an external magnetic field applied is described by the Bloch equation~\cite{Glazov2018}:
\begin{equation}\label{eq:dSt}
  \frac{\d\bm S}{\d t}=(\bm\Omega_L+\bm \Omega_N)\times\bm S-\frac{\bm S}{\tau_s}.
\end{equation}
Here $\bm S$ is the spin of the electron, $\bm\Omega_L = g_e \mu_B \bm B_z/\hbar$ is the Larmor frequency with $\mu_B$ being the Bohr magneton, $\bm\Omega_N$ is the spin precession frequency in the random Overhauser field of the nuclear spins, and $\tau_s$ is the additional phenomenological electron spin relaxation time related, for example, to the electron-phonon and spin-orbit interactions~\cite{Khaetskii2001,Woods2002}. The time scale of the nuclear spin dynamics is much longer than that of the electrons, so that $\bm\Omega_N$ can be assumed to be ``frozen''~\cite{PRBMerkulov02} and described by the probability distribution function $\mathcal F(\bm\Omega_N)\propto\exp\left[-2(\Omega_NT_2^*)^2\right]$~\cite{Hackmann2014}. Due to the pump helicity modulation in the experiment, we neglect dynamic nuclear spin polarization.

The localized electron spins are pumped and probed optically through resonant spin-singlet trion excitation, with the probe pulses arriving in effect shortly before the pump pulses. In the limit of a short trion lifetime and neglecting the electron spin dynamics between the probe and pump pulses~\cite{Smirnov2017}, following the approach of Ref.~\cite{Yugova2009} we find that the electron spin vectors before ($\bm S^-$) and after ($\bm S^+$) a pulse pair are related by:
  \begin{equation}\label{eq:cond}
    S_x^+=q^2QS_x^-,
    \quad
    S_y^+=q^2QS_y^-,
    \quad
    S_z^+=S_z^-+g.
  \end{equation}
Here $g$ is the spin polarization created by the pump pulse, $q=\cos\left(\pi/2\sqrt{P_{\text{pr}}/{P_\pi}}\right)<1$ describes the back action of the probe pulse~\cite{Zhukov2014} with $P_\pi$ indicating the $\pi$ pulse power~\cite{PRRSchering19}. The back action of the pump pulses is similarly described by $Q$~\cite{Yugova2009} \addDima{(the difference in the power of $Q$ is related to the circular polarization instead of linear)}. In agreement with the general quantum mechanical postulates~\cite{Bohr1928,braginsky1995quantum}, the measurement of the $z$ spin component destroys the transverse spin components $S_x$ and $S_y$. Microscopically, this is related to the trion excitation, which obeys the selection rules for circular polarization~\cite{Ivchenko2005} and erases the quantum coherence between the spin-up and spin-down electron states~\cite{Zhukov2010}. In fact, $1-q^2$ is the probability of trion excitation by the probe pulse. We assume the pump and probe pulses to be exactly resonant with the optical transition, which maximizes the Faraday ellipticity signal and suppresses pulse-induced spin rotations~\cite{Yugova2009}.

The steady state solution of Eqs.~\eqref{eq:dSt} and \eqref{eq:cond} yields the electron spin polarization $S_z^-$ measured by the probe pulses~\cite{supp}. This solution has to be averaged over the random nuclear fields to obtain the observed spin signal $\expval{S_z}$. Generally, this can be done only numerically~\cite{supp}, but in some limiting cases, we find transparent analytical expressions, which we give and discuss below.

A strong enough longitudinal magnetic field ($\Omega_LT_2^*\gg 1$) decouples the electron and nuclear spins. The spin relaxation, in this case, is Markovian with the time $\tau_s$, so that the measurement back action is negligible. Assuming slow spin relaxation ($\tau_s\gg T_R$), we find that $\expval{S_z}$ equals to:
\begin{equation}\label{eq:S0}
  S_0 = g\tau_s/T_R.
\end{equation}
This gives the spin polarization in the absence of hyperfine interaction.

At zero magnetic field, the electron-nuclei interaction leads to spin relaxation and thus to a smaller average spin $\left<S_z\right>$. This effect determines the amplitude and width of the PRC.

In the limit of $T_R\sim T_2^*$ and $1-q\ll 1$, both relevant for the quantum anti-Zeno effect, pulsed excitation and detection can be described as continuous spin generation and relaxation due to the measurement back action:
\begin{equation}\label{eq:kin_eq}
  \frac{\d\bm S}{\d t}=\bm\Omega_N\times\bm S-\frac{\bm S}{\tau_s}-\frac{1-q^2}{T_R}(S_x\bm e_x+S_y\bm e_y)+g\bm e_z.
\end{equation}
Here $\bm e_\alpha$ with $\alpha=x,y,z$ denote the unit vectors along the corresponding axes and the pump pulses are assumed to be weak ($Q\to 1$).

For very weak probe pulses, $q\to 1$, the third term in Eq.~\eqref{eq:kin_eq} vanishes and the average steady state solution gives~\cite{Glazov2018,PRBSmirnov20}:
\begin{equation}\label{eq:S0/3}
  \expval{S_z} = S_0/3,
\end{equation}
as expected for the isotropic hyperfine interaction~\cite{PRBMerkulov02} and observed in the experiment [Fig.~\ref{fig:1}(a)].

For finite probe pulse power $1-q$, the steady state solution of Eq.~\eqref{eq:kin_eq} can be written as $S_z=g\tau_s^{\text{eff}}/T_R$ with the effective spin relaxation time:
\begin{equation}
  \label{eq:tau_anti-Zeno}
  \tau_s^{\text{eff}}=\frac{\tau_sT_R\cos^2\theta}{T_R+(1-q^2)\tau_s\sin^2\theta},
\end{equation}
where $\theta$ is the angle between $\bm\Omega_N$ and $\bm e_z$. Analytical averaging over the nuclear fields gives:
\begin{equation}\label{eq:S_z_AZeno}
  \frac{\expval{S_z}}{S_0} =\mu\qty[\sqrt{\mu+1}\,\text{arctanh}\qty(\sqrt{1\over \mu+1})-1],
\end{equation}
where $\mu=T_R/[\tau_s(1-q^2)]$. \addDima{Since $T_R\ll\tau_s$, one has $\mu\sim1$ at $1-q\ll1$.} In the limit of $q \to 1$ this expression gives Eq.~\eqref{eq:S0/3}.

From Eq.~\eqref{eq:tau_anti-Zeno}, one can see that an increase of the probe power (decrease of $q$) leads to a decrease of the spin relaxation time $\tau_s^{\text{eff}}$ and the spin polarization $\expval{S_z}$. This is shown by the orange curve in Fig.~\ref{fig:1}(b), calculated numerically using the parameters $T_2^* = 2.2$\,ns, $\tau_s=0.5\,\mu$s, $P_\pi=2.5$\,W. We note the good agreement with the experimental results as well as with previous estimates of the system parameters~\cite{Evers2021,PRBZhukov18}.

The qualitative explanation of the quantum anti-Zeno effect is sketched at the bottom of Fig.~\ref{fig:1}. The pump pulse orients the spin polarization along $z$-axis, (i). Due to the strong Overhauser field ($T_2^*\approx T_R$), the spin polarization becomes dephased and projected onto the Overhauser field direction ($\Omega_N$) within the time between the laser pulses, (ii). After the dephasing, the action of the probe pulse projects the spin onto the $z$ axis by canceling the transverse spin components, (iii). Depending on the probe power, the cancellation varies, accelerating the spin relaxation and reducing the relative average spin polarization for higher powers at zero magnetic fields.

We highlight that the acceleration of the spin relaxation is not related to a heating effect. A higher temperature changes the electron spin relaxation time $\tau_s$, which is not related to the nuclei, only and scales the whole PRC without changing its normalized amplitude. Thus its suppression unequivocally reveals the quantum anti-Zeno effect for the nuclei-induced spin relaxation.

Moreover, the external magnetic field can be included in the numerical calculations straightforwardly, which allows us to calculate the PRC, i.e., the dependence of $\left<S_z\right>$ on $\Omega_L$. The amplitude and HWHM of the PRCs agree well with the experimental data, as shown in Figs.~\ref{fig:1}(b,c). Note that the PRC shape is not precisely Lorentzian, so the HWHM increase with increasing probe power in Fig.~\ref{fig:1}(c) can be described only qualitatively.

Next, we turn to the quantum Zeno effect, which is obtained for a long spin dephasing time, $T_2^*\gg T_R$. In this limit, from Eq.~\eqref{eq:dSt}, one can see that between the pulses, the in-plane spin components $S_{x/y}$ are increased by $\pm\Omega_{N,y/x}T_RS_z$. On the other hand, according to Eq.~\eqref{eq:cond}, the probe and pump pulses reduce them by $(1-q^2Q)S_{x/y}^-$. Thus, in the steady state, $S_{x/y}^+=\pm\Omega_{N,y/x}T_RS_z q^2Q/(1-q^2Q)$. At the same time, from Eq.~\eqref{eq:dSt} we find that the $z$ component of spin decreases between the pulses by:
\begin{multline}\label{eq:Sz_Zeno}
  S_z^+-S_z^-=\left[\frac{S_z^+}{\tau_s}+\frac{(\Omega_{N,x}^2+\Omega_{N,y}^2)T_RS_z^+}{2}\right.\\\left.+\Omega_{N,y}S_x^+-\Omega_{N,x}S_y^+\right]T_R,
\end{multline}
which is much smaller than $S_0$. This means that the spin dynamics can be described by the effective spin relaxation time $\tau_s^{\text{eff}}$, which is given by:
\begin{equation}
  \label{eq:tau_Zeno}
  \frac{1}{\tau_s^{\text{eff}}}=\frac{1}{\tau_s}+\frac{(\Omega_{N,x}^2+\Omega_{N,y}^2)T_R(1+q^2Q)}{2(1-q^2Q)}.
\end{equation}
The steady-state spin polarization equivalent to Eq.~\eqref{eq:S0} is given $S_z^-=g\tau_s^{\text{eff}}/T_R$. The averaging over the random nuclear fields can be performed analytically with the result:
\begin{equation}\label{eq:S_z_Zeno}
  \frac{\left<S_z\right>}{S_0}=-\nu{\rm{Ei}}(-\nu)\exp(\nu),
\end{equation}
where $\nu=4T_2^{*2}(1-q^2Q)/[\tau_sT_R(1+q^2Q)]$, and ${\rm{Ei}}(x)=-\int_{-x}^\infty\e^{-t}/t\d t$ is the exponential integral function. This derivation is valid for moderately strong pulses and short pulse repetition periods when $\Omega_NT_R\ll1-q^2Q$.

From Eq.~\eqref{eq:tau_Zeno}, one can see that an increase of the probe power (decrease of $q$) leads to an increase of the spin relaxation time $\tau_s^{\text{eff}}$, which is the quantum Zeno effect. This is shown by the orange curve in Fig.~\ref{fig:2}(b), calculated numerically with the parameters $T_2^*=88$\,ns, $\tau_s=0.4\,\mu$s, $P_\pi=80$\,mW, $\addDima{1-Q=0.0021}$, $g_e = -0.57$, and describes the experiment quite well, see Ref.~\cite{supp} for the $\tau_s$ and Ref.~\cite{PRBClara22} for $T_2^*$ measurements.

Qualitatively, the quantum Zeno effect can be explained as follows. Similarly, the pump pulse orients the spin polarization along the $z$-axis (i). In this case, the $T_2^* \gg T_R$. After the initialization, the spin starts to precess around the Overhauser field $\Omega_N$ (ii). For weak probe power, the effect of the probe on the transverse spin components is negligible, and the average spin polarization orients itself in the Overhauser field, leading to low spin polarization values. Once the probe power increases, it has a stronger impact on canceling the transverse spin components, as sketched at the bottom of Fig.~\ref{fig:2}, enforcing spin stabilization along the $z$ axis (iii).

\textit{Conclusion.} In this study, we have demonstrated experimentally and theoretically that the quantum measurement back action allows one to manipulate the spin relaxation time of localized electrons, which is determined by the hyperfine interaction with the nuclear spin fluctuations. The Zeno time of non-Markovian spin dynamics in the studied systems differs by more than two orders of magnitude. It allowed us to observe and describe the quantum anti-Zeno effect for quantum dots and the quantum Zeno effect for donors using the same laser repetition. The effect of measurement back action on the nuclei-induced spin relaxation was separated from the heating effect by measuring PRC by pump-probe. As an outlook we believe that this method will be successfully applied to other systems with localized electrons like colloidal quantum dots, nanoplatelets, perovskites, and Moir\'e quantum dots. Both quantum Zeno effects will be useful for operating spin-photon interfaces generating many-body entangled photon states for quantum information processing.

\textit{Acknowledgments.} We acknowledge \addDima{\href{http://www.researcherid.com/rid/E-6071-2014}{S.~A. Tarasenko}}, D.~R. Yakovlev, and N.~E. Kopteva for fruitful discussions. The financial support is provided by the Deutsche Forschungsgemeinschaft in the frame of the International Collaborative Research Center TRR 160 (Project A1). 
N.V.L and D.S.S. acknowledge the Foundation for the Advancement of Theoretical Physics and Mathematics ``BASIS''. 
A.G. and M.B. acknowledge support by the BMBF project QR.X (Contract No.16KISQ011). The theoretical model development by N.V.L and D.S.S. was supported by the Russian Science Foundation Grant No. 21-72-10035. We acknowledge the quantum dot sample supply by D. Reuter and A.~D. Wieck. The epilayer sample is provided by the Resource Center "Nanophotonics" of Saint-Petersburg State University.

\renewcommand{\i}{\ifr}
\let\oldaddcontentsline\addcontentsline
\renewcommand{\addcontentsline}[3]{}

\bibliography{citations}

\begin{thebibliography}{59}%
\makeatletter
\providecommand \@ifxundefined [1]{%
 \@ifx{#1\undefined}
}%
\providecommand \@ifnum [1]{%
 \ifnum #1\expandafter \@firstoftwo
 \else \expandafter \@secondoftwo
 \fi
}%
\providecommand \@ifx [1]{%
 \ifx #1\expandafter \@firstoftwo
 \else \expandafter \@secondoftwo
 \fi
}%
\providecommand \natexlab [1]{#1}%
\providecommand \enquote  [1]{``#1''}%
\providecommand \bibnamefont  [1]{#1}%
\providecommand \bibfnamefont [1]{#1}%
\providecommand \citenamefont [1]{#1}%
\providecommand \href@noop [0]{\@secondoftwo}%
\providecommand \href [0]{\begingroup \@sanitize@url \@href}%
\providecommand \@href[1]{\@@startlink{#1}\@@href}%
\providecommand \@@href[1]{\endgroup#1\@@endlink}%
\providecommand \@sanitize@url [0]{\catcode `\\12\catcode `\$12\catcode
  `\&12\catcode `\#12\catcode `\^12\catcode `\_12\catcode `\%12\relax}%
\providecommand \@@startlink[1]{}%
\providecommand \@@endlink[0]{}%
\providecommand \url  [0]{\begingroup\@sanitize@url \@url }%
\providecommand \@url [1]{\endgroup\@href {#1}{\urlprefix }}%
\providecommand \urlprefix  [0]{URL }%
\providecommand \Eprint [0]{\href }%
\providecommand \doibase [0]{https://doi.org/}%
\providecommand \selectlanguage [0]{\@gobble}%
\providecommand \bibinfo  [0]{\@secondoftwo}%
\providecommand \bibfield  [0]{\@secondoftwo}%
\providecommand \translation [1]{[#1]}%
\providecommand \BibitemOpen [0]{}%
\providecommand \bibitemStop [0]{}%
\providecommand \bibitemNoStop [0]{.\EOS\space}%
\providecommand \EOS [0]{\spacefactor3000\relax}%
\providecommand \BibitemShut  [1]{\csname bibitem#1\endcsname}%
\let\auto@bib@innerbib\@empty
\bibitem [{\citenamefont {Frerichs}\ and\ \citenamefont
  {Schenzle}(1991)}]{Frerichs1991}%
  \BibitemOpen
  \bibfield  {author} {\bibinfo {author} {\bibfnamefont {V.}~\bibnamefont
  {Frerichs}}\ and\ \bibinfo {author} {\bibfnamefont {A.}~\bibnamefont
  {Schenzle}},\ }\bibfield  {title} {\bibinfo {title} {{Quantum Zeno effect
  without collapse of the wave packet}},\ }\href
  {https://doi.org/10.1103/PhysRevA.44.1962} {\bibfield  {journal} {\bibinfo
  {journal} {Phys. Rev. A}\ }\textbf {\bibinfo {volume} {44}},\ \bibinfo
  {pages} {1962} (\bibinfo {year} {1991})}\BibitemShut {NoStop}%
\bibitem [{\citenamefont {Facchi}\ and\ \citenamefont
  {Pascazio}(2001)}]{Facchi2001}%
  \BibitemOpen
  \bibfield  {author} {\bibinfo {author} {\bibfnamefont {P.}~\bibnamefont
  {Facchi}}\ and\ \bibinfo {author} {\bibfnamefont {S.}~\bibnamefont
  {Pascazio}},\ }\bibfield  {title} {\bibinfo {title} {{Quantum Zeno and
  inverse quantum Zeno effects}},\ }in\ \href
  {https://doi.org/10.1016/s0079-6638(01)80017-2} {\emph {\bibinfo {booktitle}
  {{Progress in Optics}}}}\ (\bibinfo  {publisher} {Elsevier},\ \bibinfo {year}
  {2001})\ p.\ \bibinfo {pages} {147}\BibitemShut {NoStop}%
\bibitem [{\citenamefont {Kraus}(1983)}]{Kraus1983}%
  \BibitemOpen
  \bibfield  {author} {\bibinfo {author} {\bibfnamefont {K.}~\bibnamefont
  {Kraus}},\ }\href@noop {} {\bibinfo {title} {States, effects and operations,
  vol. 190 of lecture notes in physics}} (\bibinfo {year} {1983})\BibitemShut
  {NoStop}%
\bibitem [{\citenamefont {Braginsky}\ and\ \citenamefont
  {Khalili}(1996)}]{Braginsky1996}%
  \BibitemOpen
  \bibfield  {author} {\bibinfo {author} {\bibfnamefont {V.~B.}\ \bibnamefont
  {Braginsky}}\ and\ \bibinfo {author} {\bibfnamefont {F.~Y.}\ \bibnamefont
  {Khalili}},\ }\bibfield  {title} {\bibinfo {title} {Quantum nondemolition
  measurements: the route from toys to tools},\ }\href
  {https://doi.org/10.1103/RevModPhys.68.1} {\bibfield  {journal} {\bibinfo
  {journal} {Rev. Mod. Phys.}\ }\textbf {\bibinfo {volume} {68}},\ \bibinfo
  {pages} {1} (\bibinfo {year} {1996})}\BibitemShut {NoStop}%
\bibitem [{\citenamefont {Haroche}\ and\ \citenamefont
  {Raimond}(2006)}]{Haroche2006}%
  \BibitemOpen
  \bibfield  {author} {\bibinfo {author} {\bibfnamefont {S.}~\bibnamefont
  {Haroche}}\ and\ \bibinfo {author} {\bibfnamefont {J.-M.}\ \bibnamefont
  {Raimond}},\ }\href@noop {} {\emph {\bibinfo {title} {Exploring the quantum:
  atoms, cavities, and photons}}}\ (\bibinfo  {publisher} {Oxford university
  press},\ \bibinfo {year} {2006})\BibitemShut {NoStop}%
\bibitem [{\citenamefont {Clerk}\ \emph {et~al.}(2010)\citenamefont {Clerk},
  \citenamefont {Devoret}, \citenamefont {Girvin}, \citenamefont {Marquardt},\
  and\ \citenamefont {Schoelkopf}}]{Clerk2010}%
  \BibitemOpen
  \bibfield  {author} {\bibinfo {author} {\bibfnamefont {A.~A.}\ \bibnamefont
  {Clerk}}, \bibinfo {author} {\bibfnamefont {M.~H.}\ \bibnamefont {Devoret}},
  \bibinfo {author} {\bibfnamefont {S.~M.}\ \bibnamefont {Girvin}}, \bibinfo
  {author} {\bibfnamefont {F.}~\bibnamefont {Marquardt}},\ and\ \bibinfo
  {author} {\bibfnamefont {R.~J.}\ \bibnamefont {Schoelkopf}},\ }\bibfield
  {title} {\bibinfo {title} {Introduction to quantum noise, measurement, and
  amplification},\ }\href {https://doi.org/10.1103/RevModPhys.82.1155}
  {\bibfield  {journal} {\bibinfo  {journal} {Rev. Mod. Phys.}\ }\textbf
  {\bibinfo {volume} {82}},\ \bibinfo {pages} {1155} (\bibinfo {year}
  {2010})}\BibitemShut {NoStop}%
\bibitem [{\citenamefont {Raussendorf}\ \emph {et~al.}(2003)\citenamefont
  {Raussendorf}, \citenamefont {Browne},\ and\ \citenamefont
  {Briegel}}]{Raussendorf2003}%
  \BibitemOpen
  \bibfield  {author} {\bibinfo {author} {\bibfnamefont {R.}~\bibnamefont
  {Raussendorf}}, \bibinfo {author} {\bibfnamefont {D.~E.}\ \bibnamefont
  {Browne}},\ and\ \bibinfo {author} {\bibfnamefont {H.~J.}\ \bibnamefont
  {Briegel}},\ }\bibfield  {title} {\bibinfo {title} {Measurement-based quantum
  computation on cluster states},\ }\href
  {https://doi.org/10.1103/PhysRevA.68.022312} {\bibfield  {journal} {\bibinfo
  {journal} {Phys. Rev. A}\ }\textbf {\bibinfo {volume} {68}},\ \bibinfo
  {pages} {022312} (\bibinfo {year} {2003})}\BibitemShut {NoStop}%
\bibitem [{\citenamefont {Briegel}\ \emph {et~al.}(2009)\citenamefont
  {Briegel}, \citenamefont {Browne}, \citenamefont {D\"ur}, \citenamefont
  {Raussendorf},\ and\ \citenamefont {den Nest}}]{Briegel2009}%
  \BibitemOpen
  \bibfield  {author} {\bibinfo {author} {\bibfnamefont {H.~J.}\ \bibnamefont
  {Briegel}}, \bibinfo {author} {\bibfnamefont {D.~E.}\ \bibnamefont {Browne}},
  \bibinfo {author} {\bibfnamefont {W.}~\bibnamefont {D\"ur}}, \bibinfo
  {author} {\bibfnamefont {R.}~\bibnamefont {Raussendorf}},\ and\ \bibinfo
  {author} {\bibfnamefont {M.~V.}\ \bibnamefont {den Nest}},\ }\bibfield
  {title} {\bibinfo {title} {Measurement-based quantum computation},\ }\href
  {https://doi.org/10.1038/nphys1157} {\bibfield  {journal} {\bibinfo
  {journal} {Nat. Phys.}\ }\textbf {\bibinfo {volume} {5}},\ \bibinfo {pages}
  {19} (\bibinfo {year} {2009})}\BibitemShut {NoStop}%
\bibitem [{\citenamefont {Benjamin}\ \emph {et~al.}(2009)\citenamefont
  {Benjamin}, \citenamefont {Lovett},\ and\ \citenamefont
  {Smith}}]{Benjamin2009}%
  \BibitemOpen
  \bibfield  {author} {\bibinfo {author} {\bibfnamefont {S.~C.}\ \bibnamefont
  {Benjamin}}, \bibinfo {author} {\bibfnamefont {B.~W.}\ \bibnamefont
  {Lovett}},\ and\ \bibinfo {author} {\bibfnamefont {J.~M.}\ \bibnamefont
  {Smith}},\ }\bibfield  {title} {\bibinfo {title} {Prospects for
  measurement-based quantum computing with solid state spins},\ }\href
  {https://doi.org/https://doi.org/10.1002/lpor.200810051} {\bibfield
  {journal} {\bibinfo  {journal} {Laser Photon. Rev.}\ }\textbf {\bibinfo
  {volume} {3}},\ \bibinfo {pages} {556} (\bibinfo {year} {2009})}\BibitemShut
  {NoStop}%
\bibitem [{\citenamefont {Khalfin}(1958)}]{Khalfin1958}%
  \BibitemOpen
  \bibfield  {author} {\bibinfo {author} {\bibfnamefont {L.~A.}\ \bibnamefont
  {Khalfin}},\ }\bibfield  {title} {\bibinfo {title} {Contribution to the decay
  theory of a quasi-stationary state},\ }\href@noop {} {\bibfield  {journal}
  {\bibinfo  {journal} {Sov. Phys. JETP}\ }\textbf {\bibinfo {volume} {6}},\
  \bibinfo {pages} {1053} (\bibinfo {year} {1958})}\BibitemShut {NoStop}%
\bibitem [{\citenamefont {Misra}\ and\ \citenamefont
  {Sudarshan}(1977)}]{Misra1977}%
  \BibitemOpen
  \bibfield  {author} {\bibinfo {author} {\bibfnamefont {B.}~\bibnamefont
  {Misra}}\ and\ \bibinfo {author} {\bibfnamefont {E.~C.~G.}\ \bibnamefont
  {Sudarshan}},\ }\bibfield  {title} {\bibinfo {title} {The {{Zeno}}'s paradox
  in quantum theory},\ }\href {https://doi.org/10.1063/1.523304} {\bibfield
  {journal} {\bibinfo  {journal} {J. Math. Phys.}\ }\textbf {\bibinfo {volume}
  {18}},\ \bibinfo {pages} {756} (\bibinfo {year} {1977})}\BibitemShut
  {NoStop}%
\bibitem [{\citenamefont {Facchi}\ and\ \citenamefont
  {Pascazio}(2008)}]{facchi_2008}%
  \BibitemOpen
  \bibfield  {author} {\bibinfo {author} {\bibfnamefont {P.}~\bibnamefont
  {Facchi}}\ and\ \bibinfo {author} {\bibfnamefont {S.}~\bibnamefont
  {Pascazio}},\ }\bibfield  {title} {\bibinfo {title} {Quantum {{Zeno}}
  dynamics: Mathematical and physical aspects},\ }\href
  {https://doi.org/10.1088/1751-8113/41/49/493001} {\bibfield  {journal}
  {\bibinfo  {journal} {J. Phys. A: Math. Theor.}\ }\textbf {\bibinfo {volume}
  {41}},\ \bibinfo {pages} {493001} (\bibinfo {year} {2008})}\BibitemShut
  {NoStop}%
\bibitem [{\citenamefont {Kaulakys}\ and\ \citenamefont
  {Gontis}(1997)}]{Kaulakys1997}%
  \BibitemOpen
  \bibfield  {author} {\bibinfo {author} {\bibfnamefont {B.}~\bibnamefont
  {Kaulakys}}\ and\ \bibinfo {author} {\bibfnamefont {V.}~\bibnamefont
  {Gontis}},\ }\bibfield  {title} {\bibinfo {title} {{Quantum anti-Zeno
  effect}},\ }\href {https://doi.org/10.1103/PhysRevA.56.1131} {\bibfield
  {journal} {\bibinfo  {journal} {Phys. Rev. A}\ }\textbf {\bibinfo {volume}
  {56}},\ \bibinfo {pages} {1131} (\bibinfo {year} {1997})}\BibitemShut
  {NoStop}%
\bibitem [{\citenamefont {Facchi}\ \emph {et~al.}(2001)\citenamefont {Facchi},
  \citenamefont {Nakazato},\ and\ \citenamefont {Pascazio}}]{Facchi2001a}%
  \BibitemOpen
  \bibfield  {author} {\bibinfo {author} {\bibfnamefont {P.}~\bibnamefont
  {Facchi}}, \bibinfo {author} {\bibfnamefont {H.}~\bibnamefont {Nakazato}},\
  and\ \bibinfo {author} {\bibfnamefont {S.}~\bibnamefont {Pascazio}},\
  }\bibfield  {title} {\bibinfo {title} {{From the Quantum Zeno to the Inverse
  Quantum Zeno Effect}},\ }\href {https://doi.org/10.1103/PhysRevLett.86.2699}
  {\bibfield  {journal} {\bibinfo  {journal} {Phys. Rev. Lett.}\ }\textbf
  {\bibinfo {volume} {86}},\ \bibinfo {pages} {2699} (\bibinfo {year}
  {2001})}\BibitemShut {NoStop}%
\bibitem [{\citenamefont {Maniscalco}\ \emph {et~al.}(2006)\citenamefont
  {Maniscalco}, \citenamefont {Piilo},\ and\ \citenamefont
  {Suominen}}]{Maniscalco2006}%
  \BibitemOpen
  \bibfield  {author} {\bibinfo {author} {\bibfnamefont {S.}~\bibnamefont
  {Maniscalco}}, \bibinfo {author} {\bibfnamefont {J.}~\bibnamefont {Piilo}},\
  and\ \bibinfo {author} {\bibfnamefont {K.-A.}\ \bibnamefont {Suominen}},\
  }\bibfield  {title} {\bibinfo {title} {{Zeno and Anti-Zeno Effects for
  Quantum Brownian Motion}},\ }\href
  {https://doi.org/10.1103/PhysRevLett.97.130402} {\bibfield  {journal}
  {\bibinfo  {journal} {Phys. Rev. Lett.}\ }\textbf {\bibinfo {volume} {97}},\
  \bibinfo {pages} {130402} (\bibinfo {year} {2006})}\BibitemShut {NoStop}%
\bibitem [{\citenamefont {Franson}\ \emph {et~al.}(2004)\citenamefont
  {Franson}, \citenamefont {Jacobs},\ and\ \citenamefont
  {Pittman}}]{FransonQZEGate2004}%
  \BibitemOpen
  \bibfield  {author} {\bibinfo {author} {\bibfnamefont {J.~D.}\ \bibnamefont
  {Franson}}, \bibinfo {author} {\bibfnamefont {B.~C.}\ \bibnamefont
  {Jacobs}},\ and\ \bibinfo {author} {\bibfnamefont {T.~B.}\ \bibnamefont
  {Pittman}},\ }\bibfield  {title} {\bibinfo {title} {Quantum computing using
  single photons and the zeno effect},\ }\href
  {https://doi.org/10.1103/PhysRevA.70.062302} {\bibfield  {journal} {\bibinfo
  {journal} {Phys. Rev. A}\ }\textbf {\bibinfo {volume} {70}},\ \bibinfo
  {pages} {062302} (\bibinfo {year} {2004})}\BibitemShut {NoStop}%
\bibitem [{\citenamefont {Huang}\ and\ \citenamefont
  {Moore}(2008)}]{HuangQZEGate2008}%
  \BibitemOpen
  \bibfield  {author} {\bibinfo {author} {\bibfnamefont {Y.~P.}\ \bibnamefont
  {Huang}}\ and\ \bibinfo {author} {\bibfnamefont {M.~G.}\ \bibnamefont
  {Moore}},\ }\bibfield  {title} {\bibinfo {title} {Interaction- and
  measurement-free quantum zeno gates for universal computation with
  single-atom and single-photon qubits},\ }\href
  {https://doi.org/10.1103/PhysRevA.77.062332} {\bibfield  {journal} {\bibinfo
  {journal} {Phys. Rev. A}\ }\textbf {\bibinfo {volume} {77}},\ \bibinfo
  {pages} {062332} (\bibinfo {year} {2008})}\BibitemShut {NoStop}%
\bibitem [{\citenamefont {Dyakonov}(2017)}]{Dyakonov2017}%
  \BibitemOpen
  \bibinfo {editor} {\bibfnamefont {M.~I.}\ \bibnamefont {Dyakonov}},\ ed.,\
  \href@noop {} {\emph {\bibinfo {title} {Spin physics in semiconductors}}}\
  (\bibinfo  {publisher} {Springer International Publishing AG, Berlin},\
  \bibinfo {year} {2017})\BibitemShut {NoStop}%
\bibitem [{\citenamefont {Elliott}(1954)}]{Elliott1954}%
  \BibitemOpen
  \bibfield  {author} {\bibinfo {author} {\bibfnamefont {R.~J.}\ \bibnamefont
  {Elliott}},\ }\bibfield  {title} {\bibinfo {title} {{Theory of the Effect of
  Spin-Orbit Coupling on Magnetic Resonance in Some Semiconductors}},\ }\href
  {https://doi.org/10.1103/PhysRev.96.266} {\bibfield  {journal} {\bibinfo
  {journal} {Phys. Rev.}\ }\textbf {\bibinfo {volume} {96}},\ \bibinfo {pages}
  {266} (\bibinfo {year} {1954})}\BibitemShut {NoStop}%
\bibitem [{\citenamefont {Yafet}(1963)}]{Yafet1963}%
  \BibitemOpen
  \bibfield  {author} {\bibinfo {author} {\bibfnamefont {Y.}~\bibnamefont
  {Yafet}},\ }\bibinfo {title} {$g$-factors and spin-lattice relaxation of
  conduction electrons},\ in\ \href@noop {} {\emph {\bibinfo {booktitle} {Solid
  State Physics}}},\ \bibinfo {editor} {edited by\ \bibinfo {editor}
  {\bibfnamefont {F.}~\bibnamefont {Seitz}}\ and\ \bibinfo {editor}
  {\bibfnamefont {D.}~\bibnamefont {Turnbull}}}\ (\bibinfo  {publisher}
  {Academic, New-York},\ \bibinfo {year} {1963})\ p.~\bibinfo {pages}
  {2}\BibitemShut {NoStop}%
\bibitem [{\citenamefont {Bir}\ \emph {et~al.}(1975)\citenamefont {Bir},
  \citenamefont {Aronov},\ and\ \citenamefont {Pikus}}]{Bir1975}%
  \BibitemOpen
  \bibfield  {author} {\bibinfo {author} {\bibfnamefont {G.~L.}\ \bibnamefont
  {Bir}}, \bibinfo {author} {\bibfnamefont {A.~G.}\ \bibnamefont {Aronov}},\
  and\ \bibinfo {author} {\bibfnamefont {G.~E.}\ \bibnamefont {Pikus}},\
  }\bibfield  {title} {\bibinfo {title} {Spin relaxation of electrons due to
  scattering by holes},\ }\href@noop {} {\bibfield  {journal} {\bibinfo
  {journal} {Sov. Phys. JETP}\ }\textbf {\bibinfo {volume} {42}},\ \bibinfo
  {pages} {705} (\bibinfo {year} {1975})}\BibitemShut {NoStop}%
\bibitem [{\citenamefont {Dyakonov}\ and\ \citenamefont
  {Perel'}(1972)}]{Dyakonov1972}%
  \BibitemOpen
  \bibfield  {author} {\bibinfo {author} {\bibfnamefont {M.}~\bibnamefont
  {Dyakonov}}\ and\ \bibinfo {author} {\bibfnamefont {V.}~\bibnamefont
  {Perel'}},\ }\bibfield  {title} {\bibinfo {title} {Spin relaxation of
  conduction electrons in noncentrosymmetric semiconductors},\ }\href@noop {}
  {\bibfield  {journal} {\bibinfo  {journal} {Sov. Phys. Solid State}\ }\textbf
  {\bibinfo {volume} {13}},\ \bibinfo {pages} {3023} (\bibinfo {year}
  {1972})}\BibitemShut {NoStop}%
\bibitem [{\citenamefont {Merkulov}\ \emph {et~al.}(2002)\citenamefont
  {Merkulov}, \citenamefont {Efros},\ and\ \citenamefont
  {Rosen}}]{PRBMerkulov02}%
  \BibitemOpen
  \bibfield  {author} {\bibinfo {author} {\bibfnamefont {I.~A.}\ \bibnamefont
  {Merkulov}}, \bibinfo {author} {\bibfnamefont {A.~L.}\ \bibnamefont
  {Efros}},\ and\ \bibinfo {author} {\bibfnamefont {M.}~\bibnamefont {Rosen}},\
  }\bibfield  {title} {\bibinfo {title} {Electron spin relaxation by nuclei in
  semiconductor quantum dots},\ }\href
  {https://doi.org/10.1103/PhysRevB.65.205309} {\bibfield  {journal} {\bibinfo
  {journal} {Phys. Rev. B}\ }\textbf {\bibinfo {volume} {65}},\ \bibinfo
  {pages} {205309} (\bibinfo {year} {2002})}\BibitemShut {NoStop}%
\bibitem [{\citenamefont {Khaetskii}\ \emph {et~al.}(2002)\citenamefont
  {Khaetskii}, \citenamefont {Loss},\ and\ \citenamefont
  {Glazman}}]{Khaetskii2002}%
  \BibitemOpen
  \bibfield  {author} {\bibinfo {author} {\bibfnamefont {A.~V.}\ \bibnamefont
  {Khaetskii}}, \bibinfo {author} {\bibfnamefont {D.}~\bibnamefont {Loss}},\
  and\ \bibinfo {author} {\bibfnamefont {L.}~\bibnamefont {Glazman}},\
  }\bibfield  {title} {\bibinfo {title} {{Electron Spin Decoherence in Quantum
  Dots due to Interaction with Nuclei}},\ }\href
  {https://doi.org/10.1103/PhysRevLett.88.186802} {\bibfield  {journal}
  {\bibinfo  {journal} {Phys. Rev. Lett.}\ }\textbf {\bibinfo {volume} {88}},\
  \bibinfo {pages} {186802} (\bibinfo {year} {2002})}\BibitemShut {NoStop}%
\bibitem [{\citenamefont {Coish}\ and\ \citenamefont {Loss}(2004)}]{Coish2004}%
  \BibitemOpen
  \bibfield  {author} {\bibinfo {author} {\bibfnamefont {W.~A.}\ \bibnamefont
  {Coish}}\ and\ \bibinfo {author} {\bibfnamefont {D.}~\bibnamefont {Loss}},\
  }\bibfield  {title} {\bibinfo {title} {Hyperfine interaction in a quantum
  dot: Non-markovian electron spin dynamics},\ }\href
  {https://doi.org/10.1103/PhysRevB.70.195340} {\bibfield  {journal} {\bibinfo
  {journal} {Phys. Rev. B}\ }\textbf {\bibinfo {volume} {70}},\ \bibinfo
  {pages} {195340} (\bibinfo {year} {2004})}\BibitemShut {NoStop}%
\bibitem [{\citenamefont {Coish}\ and\ \citenamefont
  {Baugh}(2009)}]{Coish2009}%
  \BibitemOpen
  \bibfield  {author} {\bibinfo {author} {\bibfnamefont {W.~A.}\ \bibnamefont
  {Coish}}\ and\ \bibinfo {author} {\bibfnamefont {J.}~\bibnamefont {Baugh}},\
  }\bibfield  {title} {\bibinfo {title} {Nuclear spins in nanostructures},\
  }\href {https://doi.org/10.1002/pssb.200945229} {\bibfield  {journal}
  {\bibinfo  {journal} {Phys. Status Solidi B}\ }\textbf {\bibinfo {volume}
  {246}},\ \bibinfo {pages} {2203} (\bibinfo {year} {2009})}\BibitemShut
  {NoStop}%
\bibitem [{\citenamefont {Shumilin}\ and\ \citenamefont
  {Smirnov}(2021)}]{Shumilin2021}%
  \BibitemOpen
  \bibfield  {author} {\bibinfo {author} {\bibfnamefont {A.~V.}\ \bibnamefont
  {Shumilin}}\ and\ \bibinfo {author} {\bibfnamefont {D.~S.}\ \bibnamefont
  {Smirnov}},\ }\bibfield  {title} {\bibinfo {title} {{Nuclear Spin Dynamics,
  Noise, Squeezing, and Entanglement in Box Model}},\ }\href
  {https://doi.org/10.1103/PhysRevLett.126.216804} {\bibfield  {journal}
  {\bibinfo  {journal} {Phys. Rev. Lett.}\ }\textbf {\bibinfo {volume} {126}},\
  \bibinfo {pages} {216804} (\bibinfo {year} {2021})}\BibitemShut {NoStop}%
\bibitem [{\citenamefont {Poltavtsev}\ \emph {et~al.}(2014)\citenamefont
  {Poltavtsev}, \citenamefont {Ryzhov}, \citenamefont {Glazov}, \citenamefont
  {Koz\-lov}, \citenamefont {Zapasskii}, \citenamefont {Kavokin}, \citenamefont
  {Lagoudakis}, \citenamefont {Smirnov},\ and\ \citenamefont
  {Ivchenko}}]{Poltavtsev2014}%
  \BibitemOpen
  \bibfield  {author} {\bibinfo {author} {\bibfnamefont {S.~V.}\ \bibnamefont
  {Poltavtsev}}, \bibinfo {author} {\bibfnamefont {I.~I.}\ \bibnamefont
  {Ryzhov}}, \bibinfo {author} {\bibfnamefont {M.~M.}\ \bibnamefont {Glazov}},
  \bibinfo {author} {\bibfnamefont {G.~G.}\ \bibnamefont {Koz\-lov}}, \bibinfo
  {author} {\bibfnamefont {V.~S.}\ \bibnamefont {Zapasskii}}, \bibinfo {author}
  {\bibfnamefont {A.~V.}\ \bibnamefont {Kavokin}}, \bibinfo {author}
  {\bibfnamefont {P.~G.}\ \bibnamefont {Lagoudakis}}, \bibinfo {author}
  {\bibfnamefont {D.~S.}\ \bibnamefont {Smirnov}},\ and\ \bibinfo {author}
  {\bibfnamefont {E.~L.}\ \bibnamefont {Ivchenko}},\ }\bibfield  {title}
  {\bibinfo {title} {Spin noise spectroscopy of a single quantum well
  microcavity},\ }\href {https://doi.org/10.1103/PhysRevB.89.081304} {\bibfield
   {journal} {\bibinfo  {journal} {Phys. Rev. B}\ }\textbf {\bibinfo {volume}
  {89}},\ \bibinfo {pages} {081304} (\bibinfo {year} {2014})}\BibitemShut
  {NoStop}%
\bibitem [{\citenamefont {Leppenen}\ \emph {et~al.}(2021)\citenamefont
  {Leppenen}, \citenamefont {Lanco},\ and\ \citenamefont {Smirnov}}]{ZenoPRB}%
  \BibitemOpen
  \bibfield  {author} {\bibinfo {author} {\bibfnamefont {N.~V.}\ \bibnamefont
  {Leppenen}}, \bibinfo {author} {\bibfnamefont {L.}~\bibnamefont {Lanco}},\
  and\ \bibinfo {author} {\bibfnamefont {D.~S.}\ \bibnamefont {Smirnov}},\
  }\bibfield  {title} {\bibinfo {title} {Quantum zeno effect and quantum
  nondemolition spin measurement in a quantum dot--micropillar cavity in the
  strong coupling regime},\ }\href
  {https://doi.org/10.1103/PhysRevB.103.045413} {\bibfield  {journal} {\bibinfo
   {journal} {Phys. Rev. B}\ }\textbf {\bibinfo {volume} {103}},\ \bibinfo
  {pages} {045413} (\bibinfo {year} {2021})}\BibitemShut {NoStop}%
\bibitem [{\citenamefont {Leppenen}\ and\ \citenamefont
  {Smirnov}(2022)}]{Leppenen2022}%
  \BibitemOpen
  \bibfield  {author} {\bibinfo {author} {\bibfnamefont {N.~V.}\ \bibnamefont
  {Leppenen}}\ and\ \bibinfo {author} {\bibfnamefont {D.~S.}\ \bibnamefont
  {Smirnov}},\ }\bibfield  {title} {\bibinfo {title} {{Optical measurement of
  electron spins in quantum dots: quantum Zeno effects}},\ }\href
  {https://doi.org/10.1039/d2nr01241c} {\bibfield  {journal} {\bibinfo
  {journal} {Nanoscale}\ }\textbf {\bibinfo {volume} {14}},\ \bibinfo {pages}
  {13284} (\bibinfo {year} {2022})}\BibitemShut {NoStop}%
\bibitem [{\citenamefont {Hackenbroich}\ \emph {et~al.}(1998)\citenamefont
  {Hackenbroich}, \citenamefont {Rosenow},\ and\ \citenamefont
  {Weidenm\"uller}}]{Hackenbroich1998}%
  \BibitemOpen
  \bibfield  {author} {\bibinfo {author} {\bibfnamefont {G.}~\bibnamefont
  {Hackenbroich}}, \bibinfo {author} {\bibfnamefont {B.}~\bibnamefont
  {Rosenow}},\ and\ \bibinfo {author} {\bibfnamefont {H.~A.}\ \bibnamefont
  {Weidenm\"uller}},\ }\bibfield  {title} {\bibinfo {title} {{Quantum Zeno
  Effect and Parametric Resonance in Mesoscopic Physics}},\ }\href
  {https://doi.org/10.1103/PhysRevLett.81.5896} {\bibfield  {journal} {\bibinfo
   {journal} {Phys. Rev. Lett.}\ }\textbf {\bibinfo {volume} {81}},\ \bibinfo
  {pages} {5896} (\bibinfo {year} {1998})}\BibitemShut {NoStop}%
\bibitem [{\citenamefont {Khomitsky}\ \emph {et~al.}(2012)\citenamefont
  {Khomitsky}, \citenamefont {Gulyaev},\ and\ \citenamefont
  {Sherman}}]{Khomitsky2012}%
  \BibitemOpen
  \bibfield  {author} {\bibinfo {author} {\bibfnamefont {D.~V.}\ \bibnamefont
  {Khomitsky}}, \bibinfo {author} {\bibfnamefont {L.~V.}\ \bibnamefont
  {Gulyaev}},\ and\ \bibinfo {author} {\bibfnamefont {E.~Y.}\ \bibnamefont
  {Sherman}},\ }\bibfield  {title} {\bibinfo {title} {{Spin dynamics in a
  strongly driven system: Very slow Rabi oscillations}},\ }\href
  {https://doi.org/10.1103/PhysRevB.85.125312} {\bibfield  {journal} {\bibinfo
  {journal} {Phys. Rev. B}\ }\textbf {\bibinfo {volume} {85}},\ \bibinfo
  {pages} {125312} (\bibinfo {year} {2012})}\BibitemShut {NoStop}%
\bibitem [{\citenamefont {Kang}\ \emph {et~al.}(2017)\citenamefont {Kang},
  \citenamefont {Zhang}, \citenamefont {Xu},\ and\ \citenamefont
  {Tang}}]{Kang2017}%
  \BibitemOpen
  \bibfield  {author} {\bibinfo {author} {\bibfnamefont {L.}~\bibnamefont
  {Kang}}, \bibinfo {author} {\bibfnamefont {Y.}~\bibnamefont {Zhang}},
  \bibinfo {author} {\bibfnamefont {X.}~\bibnamefont {Xu}},\ and\ \bibinfo
  {author} {\bibfnamefont {X.}~\bibnamefont {Tang}},\ }\bibfield  {title}
  {\bibinfo {title} {Quantum measurement of a double quantum dot coupled to two
  kinds of environment},\ }\href {https://doi.org/10.1103/PhysRevB.96.235417}
  {\bibfield  {journal} {\bibinfo  {journal} {Phys. Rev. B}\ }\textbf {\bibinfo
  {volume} {96}},\ \bibinfo {pages} {235417} (\bibinfo {year}
  {2017})}\BibitemShut {NoStop}%
\bibitem [{\citenamefont {Ahmadiniaz}\ \emph {et~al.}(2022)\citenamefont
  {Ahmadiniaz}, \citenamefont {Geller}, \citenamefont {K\"onig}, \citenamefont
  {Kratzer}, \citenamefont {Lorke}, \citenamefont {Schaller},\ and\
  \citenamefont {Sch\"utzhold}}]{AhmadiniazPRR}%
  \BibitemOpen
  \bibfield  {author} {\bibinfo {author} {\bibfnamefont {N.}~\bibnamefont
  {Ahmadiniaz}}, \bibinfo {author} {\bibfnamefont {M.}~\bibnamefont {Geller}},
  \bibinfo {author} {\bibfnamefont {J.}~\bibnamefont {K\"onig}}, \bibinfo
  {author} {\bibfnamefont {P.}~\bibnamefont {Kratzer}}, \bibinfo {author}
  {\bibfnamefont {A.}~\bibnamefont {Lorke}}, \bibinfo {author} {\bibfnamefont
  {G.}~\bibnamefont {Schaller}},\ and\ \bibinfo {author} {\bibfnamefont
  {R.}~\bibnamefont {Sch\"utzhold}},\ }\bibfield  {title} {\bibinfo {title}
  {Quantum zeno manipulation of quantum dots},\ }\href
  {https://doi.org/10.1103/PhysRevResearch.4.L032045} {\bibfield  {journal}
  {\bibinfo  {journal} {Phys. Rev. Research}\ }\textbf {\bibinfo {volume}
  {4}},\ \bibinfo {pages} {L032045} (\bibinfo {year} {2022})}\BibitemShut
  {NoStop}%
\bibitem [{\citenamefont {Yamaguchi}\ \emph {et~al.}(2008)\citenamefont
  {Yamaguchi}, \citenamefont {Asano},\ and\ \citenamefont
  {Noda}}]{Yamaguchi:08}%
  \BibitemOpen
  \bibfield  {author} {\bibinfo {author} {\bibfnamefont {M.}~\bibnamefont
  {Yamaguchi}}, \bibinfo {author} {\bibfnamefont {T.}~\bibnamefont {Asano}},\
  and\ \bibinfo {author} {\bibfnamefont {S.}~\bibnamefont {Noda}},\ }\bibfield
  {title} {\bibinfo {title} {Photon emission by nanocavity-enhanced quantum
  anti-zeno effect in solid-state cavity quantum-electrodynamics},\ }\href
  {https://doi.org/10.1364/OE.16.018067} {\bibfield  {journal} {\bibinfo
  {journal} {Opt. Express}\ }\textbf {\bibinfo {volume} {16}},\ \bibinfo
  {pages} {18067} (\bibinfo {year} {2008})}\BibitemShut {NoStop}%
\bibitem [{\citenamefont {Xu}\ \emph {et~al.}(2009)\citenamefont {Xu},
  \citenamefont {Huang}, \citenamefont {Moore},\ and\ \citenamefont
  {Piermarocchi}}]{XuQZEGate2009}%
  \BibitemOpen
  \bibfield  {author} {\bibinfo {author} {\bibfnamefont {K.~J.}\ \bibnamefont
  {Xu}}, \bibinfo {author} {\bibfnamefont {Y.~P.}\ \bibnamefont {Huang}},
  \bibinfo {author} {\bibfnamefont {M.~G.}\ \bibnamefont {Moore}},\ and\
  \bibinfo {author} {\bibfnamefont {C.}~\bibnamefont {Piermarocchi}},\
  }\bibfield  {title} {\bibinfo {title} {Two-qubit conditional phase gate in
  laser-excited semiconductor quantum dots using the quantum zeno effect},\
  }\href {https://doi.org/10.1103/PhysRevLett.103.037401} {\bibfield  {journal}
  {\bibinfo  {journal} {Phys. Rev. Lett.}\ }\textbf {\bibinfo {volume} {103}},\
  \bibinfo {pages} {037401} (\bibinfo {year} {2009})}\BibitemShut {NoStop}%
\bibitem [{\citenamefont {Nutz}\ \emph {et~al.}(2019)\citenamefont {Nutz},
  \citenamefont {Androvitsaneas}, \citenamefont {Young}, \citenamefont
  {Oulton},\ and\ \citenamefont {McCutcheon}}]{NutzPRA2019}%
  \BibitemOpen
  \bibfield  {author} {\bibinfo {author} {\bibfnamefont {T.}~\bibnamefont
  {Nutz}}, \bibinfo {author} {\bibfnamefont {P.}~\bibnamefont
  {Androvitsaneas}}, \bibinfo {author} {\bibfnamefont {A.}~\bibnamefont
  {Young}}, \bibinfo {author} {\bibfnamefont {R.}~\bibnamefont {Oulton}},\ and\
  \bibinfo {author} {\bibfnamefont {D.~P.~S.}\ \bibnamefont {McCutcheon}},\
  }\bibfield  {title} {\bibinfo {title} {Stabilization of an optical transition
  energy via nuclear zeno dynamics in quantum-dot--cavity systems},\ }\href
  {https://doi.org/10.1103/PhysRevA.99.053853} {\bibfield  {journal} {\bibinfo
  {journal} {Phys. Rev. A}\ }\textbf {\bibinfo {volume} {99}},\ \bibinfo
  {pages} {053853} (\bibinfo {year} {2019})}\BibitemShut {NoStop}%
\bibitem [{\citenamefont {Klauser}\ \emph {et~al.}(2008)\citenamefont
  {Klauser}, \citenamefont {Coish},\ and\ \citenamefont {Loss}}]{Klauser2008}%
  \BibitemOpen
  \bibfield  {author} {\bibinfo {author} {\bibfnamefont {D.}~\bibnamefont
  {Klauser}}, \bibinfo {author} {\bibfnamefont {W.~A.}\ \bibnamefont {Coish}},\
  and\ \bibinfo {author} {\bibfnamefont {D.}~\bibnamefont {Loss}},\ }\bibfield
  {title} {\bibinfo {title} {{Nuclear spin dynamics and Zeno effect in quantum
  dots and defect centers}},\ }\href
  {https://doi.org/10.1103/PhysRevB.78.205301} {\bibfield  {journal} {\bibinfo
  {journal} {Phys. Rev. B}\ }\textbf {\bibinfo {volume} {78}},\ \bibinfo
  {pages} {205301} (\bibinfo {year} {2008})}\BibitemShut {NoStop}%
\bibitem [{\citenamefont {M{\k{a}}dzik}\ \emph {et~al.}(2020)\citenamefont
  {M{\k{a}}dzik}, \citenamefont {Ladd}, \citenamefont {Hudson}, \citenamefont
  {Itoh}, \citenamefont {Jakob}, \citenamefont {Johnson}, \citenamefont
  {McCallum}, \citenamefont {Jamieson}, \citenamefont {Dzurak}, \citenamefont
  {Laucht},\ and\ \citenamefont {Morello}}]{mkadzik2020controllable}%
  \BibitemOpen
  \bibfield  {author} {\bibinfo {author} {\bibfnamefont {M.~T.}\ \bibnamefont
  {M{\k{a}}dzik}}, \bibinfo {author} {\bibfnamefont {T.~D.}\ \bibnamefont
  {Ladd}}, \bibinfo {author} {\bibfnamefont {F.~E.}\ \bibnamefont {Hudson}},
  \bibinfo {author} {\bibfnamefont {K.~M.}\ \bibnamefont {Itoh}}, \bibinfo
  {author} {\bibfnamefont {A.~M.}\ \bibnamefont {Jakob}}, \bibinfo {author}
  {\bibfnamefont {B.~C.}\ \bibnamefont {Johnson}}, \bibinfo {author}
  {\bibfnamefont {J.~C.}\ \bibnamefont {McCallum}}, \bibinfo {author}
  {\bibfnamefont {D.~N.}\ \bibnamefont {Jamieson}}, \bibinfo {author}
  {\bibfnamefont {A.~S.}\ \bibnamefont {Dzurak}}, \bibinfo {author}
  {\bibfnamefont {A.}~\bibnamefont {Laucht}},\ and\ \bibinfo {author}
  {\bibfnamefont {A.}~\bibnamefont {Morello}},\ }\bibfield  {title} {\bibinfo
  {title} {Controllable freezing of the nuclear spin bath in a single-atom spin
  qubit},\ }\href@noop {} {\bibfield  {journal} {\bibinfo  {journal} {Sci.
  Adv.}\ }\textbf {\bibinfo {volume} {6}},\ \bibinfo {pages} {eaba3442}
  (\bibinfo {year} {2020})}\BibitemShut {NoStop}%
\bibitem [{\citenamefont {Maimbourg}\ \emph {et~al.}(2021)\citenamefont
  {Maimbourg}, \citenamefont {Basko}, \citenamefont {Holzmann},\ and\
  \citenamefont {Rosso}}]{PhysRevLett.126.120603}%
  \BibitemOpen
  \bibfield  {author} {\bibinfo {author} {\bibfnamefont {T.}~\bibnamefont
  {Maimbourg}}, \bibinfo {author} {\bibfnamefont {D.~M.}\ \bibnamefont
  {Basko}}, \bibinfo {author} {\bibfnamefont {M.}~\bibnamefont {Holzmann}},\
  and\ \bibinfo {author} {\bibfnamefont {A.}~\bibnamefont {Rosso}},\ }\bibfield
   {title} {\bibinfo {title} {Bath-induced zeno localization in driven
  many-body quantum systems},\ }\href
  {https://doi.org/10.1103/PhysRevLett.126.120603} {\bibfield  {journal}
  {\bibinfo  {journal} {Phys. Rev. Lett.}\ }\textbf {\bibinfo {volume} {126}},\
  \bibinfo {pages} {120603} (\bibinfo {year} {2021})}\BibitemShut {NoStop}%
\bibitem [{\citenamefont {Schering}\ \emph {et~al.}(2021)\citenamefont
  {Schering}, \citenamefont {Evers}, \citenamefont {Nedelea}, \citenamefont
  {Smirnov}, \citenamefont {Zhukov}, \citenamefont {Yakovlev}, \citenamefont
  {Bayer}, \citenamefont {Uhrig},\ and\ \citenamefont
  {Greilich}}]{PRBSchering21}%
  \BibitemOpen
  \bibfield  {author} {\bibinfo {author} {\bibfnamefont {P.}~\bibnamefont
  {Schering}}, \bibinfo {author} {\bibfnamefont {E.}~\bibnamefont {Evers}},
  \bibinfo {author} {\bibfnamefont {V.}~\bibnamefont {Nedelea}}, \bibinfo
  {author} {\bibfnamefont {D.~S.}\ \bibnamefont {Smirnov}}, \bibinfo {author}
  {\bibfnamefont {E.~A.}\ \bibnamefont {Zhukov}}, \bibinfo {author}
  {\bibfnamefont {D.~R.}\ \bibnamefont {Yakovlev}}, \bibinfo {author}
  {\bibfnamefont {M.}~\bibnamefont {Bayer}}, \bibinfo {author} {\bibfnamefont
  {G.~S.}\ \bibnamefont {Uhrig}},\ and\ \bibinfo {author} {\bibfnamefont
  {A.}~\bibnamefont {Greilich}},\ }\bibfield  {title} {\bibinfo {title}
  {Resonant spin amplification in faraday geometry},\ }\href
  {https://doi.org/10.1103/PhysRevB.103.L201301} {\bibfield  {journal}
  {\bibinfo  {journal} {Phys. Rev. B}\ }\textbf {\bibinfo {volume} {103}},\
  \bibinfo {pages} {L201301} (\bibinfo {year} {2021})}\BibitemShut {NoStop}%
\bibitem [{\citenamefont {Rittmann}\ \emph {et~al.}(2022)\citenamefont
  {Rittmann}, \citenamefont {Petrov}, \citenamefont {Kamenskii}, \citenamefont
  {Kavokin}, \citenamefont {Kuntsevich}, \citenamefont {Efimov}, \citenamefont
  {Eliseev}, \citenamefont {Bayer},\ and\ \citenamefont
  {Greilich}}]{PRBClara22}%
  \BibitemOpen
  \bibfield  {author} {\bibinfo {author} {\bibfnamefont {C.}~\bibnamefont
  {Rittmann}}, \bibinfo {author} {\bibfnamefont {M.~Y.}\ \bibnamefont
  {Petrov}}, \bibinfo {author} {\bibfnamefont {A.~N.}\ \bibnamefont
  {Kamenskii}}, \bibinfo {author} {\bibfnamefont {K.~V.}\ \bibnamefont
  {Kavokin}}, \bibinfo {author} {\bibfnamefont {A.~Y.}\ \bibnamefont
  {Kuntsevich}}, \bibinfo {author} {\bibfnamefont {Y.~P.}\ \bibnamefont
  {Efimov}}, \bibinfo {author} {\bibfnamefont {S.~A.}\ \bibnamefont {Eliseev}},
  \bibinfo {author} {\bibfnamefont {M.}~\bibnamefont {Bayer}},\ and\ \bibinfo
  {author} {\bibfnamefont {A.}~\bibnamefont {Greilich}},\ }\bibfield  {title}
  {\bibinfo {title} {Unveiling the electron-nuclear spin dynamics in an
  $n$-doped ingaas epilayer by spin noise spectroscopy},\ }\href
  {https://doi.org/10.1103/PhysRevB.106.035202} {\bibfield  {journal} {\bibinfo
   {journal} {Phys. Rev. B}\ }\textbf {\bibinfo {volume} {106}},\ \bibinfo
  {pages} {035202} (\bibinfo {year} {2022})}\BibitemShut {NoStop}%
\bibitem [{\citenamefont {Glazov}(2018)}]{Glazov2018}%
  \BibitemOpen
  \bibfield  {author} {\bibinfo {author} {\bibfnamefont {M.~M.}\ \bibnamefont
  {Glazov}},\ }\href@noop {} {\emph {\bibinfo {title} {Electron and Nuclear
  Spin Dynamics in Semiconductor Nanostructures}}}\ (\bibinfo  {publisher}
  {Oxford University Press, Oxford},\ \bibinfo {year} {2018})\BibitemShut
  {NoStop}%
\bibitem [{Note1()}]{Note1}%
  \BibitemOpen
  \bibinfo {note} {The vertical offset of the values in Fig.~\ref {fig:1}(d)
  relative to panel (b) is related to the different pump power. The higher pump
  power in these measurements is a compromise between the noise level at very
  low pump and probe powers and the effect of the pump on the spin
  polarization.}\BibitemShut {Stop}%
\bibitem [{\citenamefont {Khaetskii}\ and\ \citenamefont
  {Nazarov}(2001)}]{Khaetskii2001}%
  \BibitemOpen
  \bibfield  {author} {\bibinfo {author} {\bibfnamefont {A.~V.}\ \bibnamefont
  {Khaetskii}}\ and\ \bibinfo {author} {\bibfnamefont {Y.~V.}\ \bibnamefont
  {Nazarov}},\ }\bibfield  {title} {\bibinfo {title} {Spin-flip transitions
  between {Z}eeman sublevels in semiconductor quantum dots},\ }\href
  {https://doi.org/10.1103/PhysRevB.64.125316} {\bibfield  {journal} {\bibinfo
  {journal} {Phys. Rev. B}\ }\textbf {\bibinfo {volume} {64}},\ \bibinfo
  {pages} {125316} (\bibinfo {year} {2001})}\BibitemShut {NoStop}%
\bibitem [{\citenamefont {Woods}\ \emph {et~al.}(2002)\citenamefont {Woods},
  \citenamefont {Reinecke},\ and\ \citenamefont {Lyanda-Geller}}]{Woods2002}%
  \BibitemOpen
  \bibfield  {author} {\bibinfo {author} {\bibfnamefont {L.~M.}\ \bibnamefont
  {Woods}}, \bibinfo {author} {\bibfnamefont {T.~L.}\ \bibnamefont
  {Reinecke}},\ and\ \bibinfo {author} {\bibfnamefont {Y.}~\bibnamefont
  {Lyanda-Geller}},\ }\bibfield  {title} {\bibinfo {title} {Spin relaxation in
  quantum dots},\ }\href {https://doi.org/10.1103/PhysRevB.66.161318}
  {\bibfield  {journal} {\bibinfo  {journal} {Phys. Rev. B}\ }\textbf {\bibinfo
  {volume} {66}},\ \bibinfo {pages} {161318(R)} (\bibinfo {year}
  {2002})}\BibitemShut {NoStop}%
\bibitem [{\citenamefont {Hackmann}\ \emph {et~al.}(2014)\citenamefont
  {Hackmann}, \citenamefont {Smirnov}, \citenamefont {Glazov},\ and\
  \citenamefont {Anders}}]{Hackmann2014}%
  \BibitemOpen
  \bibfield  {author} {\bibinfo {author} {\bibfnamefont {J.}~\bibnamefont
  {Hackmann}}, \bibinfo {author} {\bibfnamefont {D.~S.}\ \bibnamefont
  {Smirnov}}, \bibinfo {author} {\bibfnamefont {M.~M.}\ \bibnamefont
  {Glazov}},\ and\ \bibinfo {author} {\bibfnamefont {F.~B.}\ \bibnamefont
  {Anders}},\ }\bibfield  {title} {\bibinfo {title} {{Spin noise in a quantum
  dot ensemble: From a quantum mechanical to a semi-classical description}},\
  }\href {https://doi.org/10.1002/pssb.201451103} {\bibfield  {journal}
  {\bibinfo  {journal} {Phys. Status Solidi B}\ }\textbf {\bibinfo {volume}
  {251}},\ \bibinfo {pages} {1270} (\bibinfo {year} {2014})}\BibitemShut
  {NoStop}%
\bibitem [{\citenamefont {Smirnov}\ \emph {et~al.}(2017)\citenamefont
  {Smirnov}, \citenamefont {Glasenapp}, \citenamefont {Bergen}, \citenamefont
  {Glazov}, \citenamefont {Reuter}, \citenamefont {Wieck}, \citenamefont
  {Bayer},\ and\ \citenamefont {Greilich}}]{Smirnov2017}%
  \BibitemOpen
  \bibfield  {author} {\bibinfo {author} {\bibfnamefont {D.~S.}\ \bibnamefont
  {Smirnov}}, \bibinfo {author} {\bibfnamefont {P.}~\bibnamefont {Glasenapp}},
  \bibinfo {author} {\bibfnamefont {M.}~\bibnamefont {Bergen}}, \bibinfo
  {author} {\bibfnamefont {M.~M.}\ \bibnamefont {Glazov}}, \bibinfo {author}
  {\bibfnamefont {D.}~\bibnamefont {Reuter}}, \bibinfo {author} {\bibfnamefont
  {A.~D.}\ \bibnamefont {Wieck}}, \bibinfo {author} {\bibfnamefont
  {M.}~\bibnamefont {Bayer}},\ and\ \bibinfo {author} {\bibfnamefont
  {A.}~\bibnamefont {Greilich}},\ }\bibfield  {title} {\bibinfo {title}
  {Nonequilibrium spin noise in a quantum dot ensemble},\ }\href
  {https://doi.org/10.1103/PhysRevB.95.241408} {\bibfield  {journal} {\bibinfo
  {journal} {Phys. Rev. B}\ }\textbf {\bibinfo {volume} {95}},\ \bibinfo
  {pages} {241408} (\bibinfo {year} {2017})}\BibitemShut {NoStop}%
\bibitem [{\citenamefont {Yugova}\ \emph {et~al.}(2009)\citenamefont {Yugova},
  \citenamefont {Glazov}, \citenamefont {Ivchenko},\ and\ \citenamefont
  {Efros}}]{Yugova2009}%
  \BibitemOpen
  \bibfield  {author} {\bibinfo {author} {\bibfnamefont {I.~A.}\ \bibnamefont
  {Yugova}}, \bibinfo {author} {\bibfnamefont {M.~M.}\ \bibnamefont {Glazov}},
  \bibinfo {author} {\bibfnamefont {E.~L.}\ \bibnamefont {Ivchenko}},\ and\
  \bibinfo {author} {\bibfnamefont {A.~L.}\ \bibnamefont {Efros}},\ }\bibfield
  {title} {\bibinfo {title} {{Pump-probe Faraday rotation and ellipticity in an
  ensemble of singly charged quantum dots}},\ }\href
  {https://doi.org/10.1103/PhysRevB.80.104436} {\bibfield  {journal} {\bibinfo
  {journal} {Phys. Rev. B}\ }\textbf {\bibinfo {volume} {80}},\ \bibinfo {eid}
  {104436} (\bibinfo {year} {2009})}\BibitemShut {NoStop}%
\bibitem [{\citenamefont {Zhukov}\ \emph {et~al.}(2014)\citenamefont {Zhukov},
  \citenamefont {Greilich}, \citenamefont {Yakovlev}, \citenamefont {Kavokin},
  \citenamefont {Yugova}, \citenamefont {Yugov}, \citenamefont {Suter},
  \citenamefont {Karczewski}, \citenamefont {Wojtowicz}, \citenamefont
  {Kossut}, \citenamefont {Petrov}, \citenamefont {Dolgikh}, \citenamefont
  {Pawlis},\ and\ \citenamefont {Bayer}}]{Zhukov2014}%
  \BibitemOpen
  \bibfield  {author} {\bibinfo {author} {\bibfnamefont {E.~A.}\ \bibnamefont
  {Zhukov}}, \bibinfo {author} {\bibfnamefont {A.}~\bibnamefont {Greilich}},
  \bibinfo {author} {\bibfnamefont {D.~R.}\ \bibnamefont {Yakovlev}}, \bibinfo
  {author} {\bibfnamefont {K.~V.}\ \bibnamefont {Kavokin}}, \bibinfo {author}
  {\bibfnamefont {I.~A.}\ \bibnamefont {Yugova}}, \bibinfo {author}
  {\bibfnamefont {O.~A.}\ \bibnamefont {Yugov}}, \bibinfo {author}
  {\bibfnamefont {D.}~\bibnamefont {Suter}}, \bibinfo {author} {\bibfnamefont
  {G.}~\bibnamefont {Karczewski}}, \bibinfo {author} {\bibfnamefont
  {T.}~\bibnamefont {Wojtowicz}}, \bibinfo {author} {\bibfnamefont
  {J.}~\bibnamefont {Kossut}}, \bibinfo {author} {\bibfnamefont {V.~V.}\
  \bibnamefont {Petrov}}, \bibinfo {author} {\bibfnamefont {Y.~K.}\
  \bibnamefont {Dolgikh}}, \bibinfo {author} {\bibfnamefont {A.}~\bibnamefont
  {Pawlis}},\ and\ \bibinfo {author} {\bibfnamefont {M.}~\bibnamefont
  {Bayer}},\ }\bibfield  {title} {\bibinfo {title} {{All-optical NMR in
  semiconductors provided by resonant cooling of nuclear spins interacting with
  electrons in the resonant spin amplification regime}},\ }\href
  {https://doi.org/10.1103/PhysRevB.90.085311} {\bibfield  {journal} {\bibinfo
  {journal} {Phys. Rev. B}\ }\textbf {\bibinfo {volume} {90}},\ \bibinfo
  {pages} {085311} (\bibinfo {year} {2014})}\BibitemShut {NoStop}%
\bibitem [{\citenamefont {Schering}\ \emph {et~al.}(2019)\citenamefont
  {Schering}, \citenamefont {Uhrig},\ and\ \citenamefont
  {Smirnov}}]{PRRSchering19}%
  \BibitemOpen
  \bibfield  {author} {\bibinfo {author} {\bibfnamefont {P.}~\bibnamefont
  {Schering}}, \bibinfo {author} {\bibfnamefont {G.~S.}\ \bibnamefont
  {Uhrig}},\ and\ \bibinfo {author} {\bibfnamefont {D.~S.}\ \bibnamefont
  {Smirnov}},\ }\bibfield  {title} {\bibinfo {title} {Spin inertia and
  polarization recovery in quantum dots: Role of pumping strength and resonant
  spin amplification},\ }\href
  {https://doi.org/10.1103/PhysRevResearch.1.033189} {\bibfield  {journal}
  {\bibinfo  {journal} {Phys. Rev. Research}\ }\textbf {\bibinfo {volume}
  {1}},\ \bibinfo {pages} {033189} (\bibinfo {year} {2019})}\BibitemShut
  {NoStop}%
\bibitem [{\citenamefont {Bohr}(1928)}]{Bohr1928}%
  \BibitemOpen
  \bibfield  {author} {\bibinfo {author} {\bibfnamefont {N.}~\bibnamefont
  {Bohr}},\ }\bibfield  {title} {\bibinfo {title} {{The Quantum Postulate and
  the Recent Development of Atomic Theory}},\ }\href
  {https://doi.org/10.1038/121580a0} {\bibfield  {journal} {\bibinfo  {journal}
  {Nature}\ }\textbf {\bibinfo {volume} {121}},\ \bibinfo {pages} {580}
  (\bibinfo {year} {1928})}\BibitemShut {NoStop}%
\bibitem [{\citenamefont {Braginsky}\ and\ \citenamefont
  {Khalili}(1995)}]{braginsky1995quantum}%
  \BibitemOpen
  \bibfield  {author} {\bibinfo {author} {\bibfnamefont {V.~B.}\ \bibnamefont
  {Braginsky}}\ and\ \bibinfo {author} {\bibfnamefont {F.~Y.}\ \bibnamefont
  {Khalili}},\ }\href@noop {} {\emph {\bibinfo {title} {Quantum measurement}}}\
  (\bibinfo  {publisher} {Cambridge University Press},\ \bibinfo {year}
  {1995})\BibitemShut {NoStop}%
\bibitem [{\citenamefont {Ivchenko}(2005)}]{Ivchenko2005}%
  \BibitemOpen
  \bibfield  {author} {\bibinfo {author} {\bibfnamefont {E.~L.}\ \bibnamefont
  {Ivchenko}},\ }\href@noop {} {\emph {\bibinfo {title} {Optical spectroscopy
  of semiconductor nanostructures}}}\ (\bibinfo  {publisher} {Alpha Science,
  Harrow UK},\ \bibinfo {year} {2005})\BibitemShut {NoStop}%
\bibitem [{\citenamefont {Zhukov}\ \emph {et~al.}(2010)\citenamefont {Zhukov},
  \citenamefont {Yakovlev}, \citenamefont {Glazov}, \citenamefont {Fokina},
  \citenamefont {Karczewski}, \citenamefont {Wojtowicz}, \citenamefont
  {Kossut},\ and\ \citenamefont {Bayer}}]{Zhukov2010}%
  \BibitemOpen
  \bibfield  {author} {\bibinfo {author} {\bibfnamefont {E.~A.}\ \bibnamefont
  {Zhukov}}, \bibinfo {author} {\bibfnamefont {D.~R.}\ \bibnamefont
  {Yakovlev}}, \bibinfo {author} {\bibfnamefont {M.~M.}\ \bibnamefont
  {Glazov}}, \bibinfo {author} {\bibfnamefont {L.}~\bibnamefont {Fokina}},
  \bibinfo {author} {\bibfnamefont {G.}~\bibnamefont {Karczewski}}, \bibinfo
  {author} {\bibfnamefont {T.}~\bibnamefont {Wojtowicz}}, \bibinfo {author}
  {\bibfnamefont {J.}~\bibnamefont {Kossut}},\ and\ \bibinfo {author}
  {\bibfnamefont {M.}~\bibnamefont {Bayer}},\ }\bibfield  {title} {\bibinfo
  {title} {Optical control of electron spin coherence in
  $\mbox{CdTe/(Cd,Mg)Te}$ quantum wells},\ }\href
  {https://doi.org/10.1103/PhysRevB.81.235320} {\bibfield  {journal} {\bibinfo
  {journal} {Phys. Rev. B}\ }\textbf {\bibinfo {volume} {81}},\ \bibinfo
  {pages} {235320} (\bibinfo {year} {2010})}\BibitemShut {NoStop}%
\bibitem [{sup()}]{supp}%
  \BibitemOpen
  \href@noop {} {\bibinfo {title} {See supplemental material for the additional
  quantum anti-zeno effect measurements for a complementary qds sample,
  spin-inertia measurements for the sample s2, and additional theoretical
  details about finding the steady state solution, numerical averaging over the
  random nuclear fields and comparison of the analytical and numerical
  calculations.}}\BibitemShut {Stop}%
\bibitem [{\citenamefont {Smirnov}\ \emph {et~al.}(2020)\citenamefont
  {Smirnov}, \citenamefont {Zhukov}, \citenamefont {Yakovlev}, \citenamefont
  {Kirstein}, \citenamefont {Bayer},\ and\ \citenamefont
  {Greilich}}]{PRBSmirnov20}%
  \BibitemOpen
  \bibfield  {author} {\bibinfo {author} {\bibfnamefont {D.~S.}\ \bibnamefont
  {Smirnov}}, \bibinfo {author} {\bibfnamefont {E.~A.}\ \bibnamefont {Zhukov}},
  \bibinfo {author} {\bibfnamefont {D.~R.}\ \bibnamefont {Yakovlev}}, \bibinfo
  {author} {\bibfnamefont {E.}~\bibnamefont {Kirstein}}, \bibinfo {author}
  {\bibfnamefont {M.}~\bibnamefont {Bayer}},\ and\ \bibinfo {author}
  {\bibfnamefont {A.}~\bibnamefont {Greilich}},\ }\bibfield  {title} {\bibinfo
  {title} {Spin polarization recovery and hanle effect for charge carriers
  interacting with nuclear spins in semiconductors},\ }\href
  {https://doi.org/10.1103/PhysRevB.102.235413} {\bibfield  {journal} {\bibinfo
   {journal} {Phys. Rev. B}\ }\textbf {\bibinfo {volume} {102}},\ \bibinfo
  {pages} {235413} (\bibinfo {year} {2020})}\BibitemShut {NoStop}%
\bibitem [{\citenamefont {Evers}\ \emph {et~al.}(2021)\citenamefont {Evers},
  \citenamefont {Kopteva}, \citenamefont {Yugova}, \citenamefont {Yakovlev},
  \citenamefont {Reuter}, \citenamefont {Wieck}, \citenamefont {Bayer},\ and\
  \citenamefont {Greilich}}]{Evers2021}%
  \BibitemOpen
  \bibfield  {author} {\bibinfo {author} {\bibfnamefont {E.}~\bibnamefont
  {Evers}}, \bibinfo {author} {\bibfnamefont {N.~E.}\ \bibnamefont {Kopteva}},
  \bibinfo {author} {\bibfnamefont {I.~A.}\ \bibnamefont {Yugova}}, \bibinfo
  {author} {\bibfnamefont {D.~R.}\ \bibnamefont {Yakovlev}}, \bibinfo {author}
  {\bibfnamefont {D.}~\bibnamefont {Reuter}}, \bibinfo {author} {\bibfnamefont
  {A.~D.}\ \bibnamefont {Wieck}}, \bibinfo {author} {\bibfnamefont
  {M.}~\bibnamefont {Bayer}},\ and\ \bibinfo {author} {\bibfnamefont
  {A.}~\bibnamefont {Greilich}},\ }\bibfield  {title} {\bibinfo {title}
  {Suppression of nuclear spin fluctuations in an {InGaAs} quantum dot ensemble
  by {GHz}-pulsed optical excitation},\ }\href
  {https://doi.org/10.1038/s41534-021-00395-1} {\bibfield  {journal} {\bibinfo
  {journal} {npj Quantum Information}\ }\textbf {\bibinfo {volume} {7}},\
  \bibinfo {pages} {60} (\bibinfo {year} {2021})}\BibitemShut {NoStop}%
\bibitem [{\citenamefont {Zhukov}\ \emph {et~al.}(2018)\citenamefont {Zhukov},
  \citenamefont {Kirstein}, \citenamefont {Smirnov}, \citenamefont {Yakovlev},
  \citenamefont {Glazov}, \citenamefont {Reuter}, \citenamefont {Wieck},
  \citenamefont {Bayer},\ and\ \citenamefont {Greilich}}]{PRBZhukov18}%
  \BibitemOpen
  \bibfield  {author} {\bibinfo {author} {\bibfnamefont {E.~A.}\ \bibnamefont
  {Zhukov}}, \bibinfo {author} {\bibfnamefont {E.}~\bibnamefont {Kirstein}},
  \bibinfo {author} {\bibfnamefont {D.~S.}\ \bibnamefont {Smirnov}}, \bibinfo
  {author} {\bibfnamefont {D.~R.}\ \bibnamefont {Yakovlev}}, \bibinfo {author}
  {\bibfnamefont {M.~M.}\ \bibnamefont {Glazov}}, \bibinfo {author}
  {\bibfnamefont {D.}~\bibnamefont {Reuter}}, \bibinfo {author} {\bibfnamefont
  {A.~D.}\ \bibnamefont {Wieck}}, \bibinfo {author} {\bibfnamefont
  {M.}~\bibnamefont {Bayer}},\ and\ \bibinfo {author} {\bibfnamefont
  {A.}~\bibnamefont {Greilich}},\ }\bibfield  {title} {\bibinfo {title} {Spin
  inertia of resident and photoexcited carriers in singly charged quantum
  dots},\ }\href {https://doi.org/10.1103/PhysRevB.98.121304} {\bibfield
  {journal} {\bibinfo  {journal} {Phys. Rev. B}\ }\textbf {\bibinfo {volume}
  {98}},\ \bibinfo {pages} {121304} (\bibinfo {year} {2018})}\BibitemShut
  {NoStop}%
\end{thebibliography}%


\begin{thebibliography}{62}%
\makeatletter
\providecommand \@ifxundefined [1]{%
 \@ifx{#1\undefined}
}%
\providecommand \@ifnum [1]{%
 \ifnum #1\expandafter \@firstoftwo
 \else \expandafter \@secondoftwo
 \fi
}%
\providecommand \@ifx [1]{%
 \ifx #1\expandafter \@firstoftwo
 \else \expandafter \@secondoftwo
 \fi
}%
\providecommand \natexlab [1]{#1}%
\providecommand \enquote  [1]{``#1''}%
\providecommand \bibnamefont  [1]{#1}%
\providecommand \bibfnamefont [1]{#1}%
\providecommand \citenamefont [1]{#1}%
\providecommand \href@noop [0]{\@secondoftwo}%
\providecommand \href [0]{\begingroup \@sanitize@url \@href}%
\providecommand \@href[1]{\@@startlink{#1}\@@href}%
\providecommand \@@href[1]{\endgroup#1\@@endlink}%
\providecommand \@sanitize@url [0]{\catcode `\\12\catcode `\$12\catcode
  `\&12\catcode `\#12\catcode `\^12\catcode `\_12\catcode `\%12\relax}%
\providecommand \@@startlink[1]{}%
\providecommand \@@endlink[0]{}%
\providecommand \url  [0]{\begingroup\@sanitize@url \@url }%
\providecommand \@url [1]{\endgroup\@href {#1}{\urlprefix }}%
\providecommand \urlprefix  [0]{URL }%
\providecommand \Eprint [0]{\href }%
\providecommand \doibase [0]{https://doi.org/}%
\providecommand \selectlanguage [0]{\@gobble}%
\providecommand \bibinfo  [0]{\@secondoftwo}%
\providecommand \bibfield  [0]{\@secondoftwo}%
\providecommand \translation [1]{[#1]}%
\providecommand \BibitemOpen [0]{}%
\providecommand \bibitemStop [0]{}%
\providecommand \bibitemNoStop [0]{.\EOS\space}%
\providecommand \EOS [0]{\spacefactor3000\relax}%
\providecommand \BibitemShut  [1]{\csname bibitem#1\endcsname}%
\let\auto@bib@innerbib\@empty
\bibitem [{\citenamefont {Schering}\ \emph {et~al.}(2021)\citenamefont
  {Schering}, \citenamefont {Evers}, \citenamefont {Nedelea}, \citenamefont
  {Smirnov}, \citenamefont {Zhukov}, \citenamefont {Yakovlev}, \citenamefont
  {Bayer}, \citenamefont {Uhrig},\ and\ \citenamefont
  {Greilich}}]{S_PRBSchering21}%
  \BibitemOpen
  \bibfield  {author} {\bibinfo {author} {\bibfnamefont {P.}~\bibnamefont
  {Schering}}, \bibinfo {author} {\bibfnamefont {E.}~\bibnamefont {Evers}},
  \bibinfo {author} {\bibfnamefont {V.}~\bibnamefont {Nedelea}}, \bibinfo
  {author} {\bibfnamefont {D.~S.}\ \bibnamefont {Smirnov}}, \bibinfo {author}
  {\bibfnamefont {E.~A.}\ \bibnamefont {Zhukov}}, \bibinfo {author}
  {\bibfnamefont {D.~R.}\ \bibnamefont {Yakovlev}}, \bibinfo {author}
  {\bibfnamefont {M.}~\bibnamefont {Bayer}}, \bibinfo {author} {\bibfnamefont
  {G.~S.}\ \bibnamefont {Uhrig}},\ and\ \bibinfo {author} {\bibfnamefont
  {A.}~\bibnamefont {Greilich}},\ }\bibfield  {title} {\bibinfo {title}
  {Resonant spin amplification in faraday geometry},\ }\href
  {https://doi.org/10.1103/PhysRevB.103.L201301} {\bibfield  {journal}
  {\bibinfo  {journal} {Phys. Rev. B}\ }\textbf {\bibinfo {volume} {103}},\
  \bibinfo {pages} {L201301} (\bibinfo {year} {2021})}\BibitemShut {NoStop}%
\bibitem [{\citenamefont {Shabaev}\ \emph {et~al.}(2003)\citenamefont
  {Shabaev}, \citenamefont {Efros}, \citenamefont {Gammon},\ and\ \citenamefont
  {Merkulov}}]{S_PRBShabaev03}%
  \BibitemOpen
  \bibfield  {author} {\bibinfo {author} {\bibfnamefont {A.}~\bibnamefont
  {Shabaev}}, \bibinfo {author} {\bibfnamefont {A.~L.}\ \bibnamefont {Efros}},
  \bibinfo {author} {\bibfnamefont {D.}~\bibnamefont {Gammon}},\ and\ \bibinfo
  {author} {\bibfnamefont {I.~A.}\ \bibnamefont {Merkulov}},\ }\bibfield
  {title} {\bibinfo {title} {Optical readout and initialization of an electron
  spin in a single quantum dot},\ }\href
  {https://doi.org/10.1103/PhysRevB.68.201305} {\bibfield  {journal} {\bibinfo
  {journal} {Phys. Rev. B}\ }\textbf {\bibinfo {volume} {68}},\ \bibinfo
  {pages} {201305} (\bibinfo {year} {2003})}\BibitemShut {NoStop}%
\bibitem [{\citenamefont {Heisterkamp}\ \emph {et~al.}(2015)\citenamefont
  {Heisterkamp}, \citenamefont {Zhukov}, \citenamefont {Greilich},
  \citenamefont {Yakovlev}, \citenamefont {Korenev}, \citenamefont {Pawlis},\
  and\ \citenamefont {Bayer}}]{S_PRBHeisterkamp15}%
  \BibitemOpen
  \bibfield  {author} {\bibinfo {author} {\bibfnamefont {F.}~\bibnamefont
  {Heisterkamp}}, \bibinfo {author} {\bibfnamefont {E.~A.}\ \bibnamefont
  {Zhukov}}, \bibinfo {author} {\bibfnamefont {A.}~\bibnamefont {Greilich}},
  \bibinfo {author} {\bibfnamefont {D.~R.}\ \bibnamefont {Yakovlev}}, \bibinfo
  {author} {\bibfnamefont {V.~L.}\ \bibnamefont {Korenev}}, \bibinfo {author}
  {\bibfnamefont {A.}~\bibnamefont {Pawlis}},\ and\ \bibinfo {author}
  {\bibfnamefont {M.}~\bibnamefont {Bayer}},\ }\bibfield  {title} {\bibinfo
  {title} {Longitudinal and transverse spin dynamics of donor-bound electrons
  in fluorine-doped znse: Spin inertia versus hanle effect},\ }\href
  {https://doi.org/10.1103/PhysRevB.91.235432} {\bibfield  {journal} {\bibinfo
  {journal} {Phys. Rev. B}\ }\textbf {\bibinfo {volume} {91}},\ \bibinfo
  {pages} {235432} (\bibinfo {year} {2015})}\BibitemShut {NoStop}%
\bibitem [{\citenamefont {Zhukov}\ \emph {et~al.}(2018)\citenamefont {Zhukov},
  \citenamefont {Kirstein}, \citenamefont {Smirnov}, \citenamefont {Yakovlev},
  \citenamefont {Glazov}, \citenamefont {Reuter}, \citenamefont {Wieck},
  \citenamefont {Bayer},\ and\ \citenamefont {Greilich}}]{S_PRBZhukov18}%
  \BibitemOpen
  \bibfield  {author} {\bibinfo {author} {\bibfnamefont {E.~A.}\ \bibnamefont
  {Zhukov}}, \bibinfo {author} {\bibfnamefont {E.}~\bibnamefont {Kirstein}},
  \bibinfo {author} {\bibfnamefont {D.~S.}\ \bibnamefont {Smirnov}}, \bibinfo
  {author} {\bibfnamefont {D.~R.}\ \bibnamefont {Yakovlev}}, \bibinfo {author}
  {\bibfnamefont {M.~M.}\ \bibnamefont {Glazov}}, \bibinfo {author}
  {\bibfnamefont {D.}~\bibnamefont {Reuter}}, \bibinfo {author} {\bibfnamefont
  {A.~D.}\ \bibnamefont {Wieck}}, \bibinfo {author} {\bibfnamefont
  {M.}~\bibnamefont {Bayer}},\ and\ \bibinfo {author} {\bibfnamefont
  {A.}~\bibnamefont {Greilich}},\ }\bibfield  {title} {\bibinfo {title} {Spin
  inertia of resident and photoexcited carriers in singly charged quantum
  dots},\ }\href {https://doi.org/10.1103/PhysRevB.98.121304} {\bibfield
  {journal} {\bibinfo  {journal} {Phys. Rev. B}\ }\textbf {\bibinfo {volume}
  {98}},\ \bibinfo {pages} {121304} (\bibinfo {year} {2018})}\BibitemShut
  {NoStop}%
\bibitem [{\citenamefont {Smirnov}\ \emph {et~al.}(2018)\citenamefont
  {Smirnov}, \citenamefont {Zhukov}, \citenamefont {Kirstein}, \citenamefont
  {Yakovlev}, \citenamefont {Reuter}, \citenamefont {Wieck}, \citenamefont
  {Bayer}, \citenamefont {Greilich},\ and\ \citenamefont
  {Glazov}}]{S_PRBSmirnov18}%
  \BibitemOpen
  \bibfield  {author} {\bibinfo {author} {\bibfnamefont {D.~S.}\ \bibnamefont
  {Smirnov}}, \bibinfo {author} {\bibfnamefont {E.~A.}\ \bibnamefont {Zhukov}},
  \bibinfo {author} {\bibfnamefont {E.}~\bibnamefont {Kirstein}}, \bibinfo
  {author} {\bibfnamefont {D.~R.}\ \bibnamefont {Yakovlev}}, \bibinfo {author}
  {\bibfnamefont {D.}~\bibnamefont {Reuter}}, \bibinfo {author} {\bibfnamefont
  {A.~D.}\ \bibnamefont {Wieck}}, \bibinfo {author} {\bibfnamefont
  {M.}~\bibnamefont {Bayer}}, \bibinfo {author} {\bibfnamefont
  {A.}~\bibnamefont {Greilich}},\ and\ \bibinfo {author} {\bibfnamefont
  {M.~M.}\ \bibnamefont {Glazov}},\ }\bibfield  {title} {\bibinfo {title}
  {Theory of spin inertia in singly charged quantum dots},\ }\href
  {https://doi.org/10.1103/PhysRevB.98.125306} {\bibfield  {journal} {\bibinfo
  {journal} {Phys. Rev. B}\ }\textbf {\bibinfo {volume} {98}},\ \bibinfo
  {pages} {125306} (\bibinfo {year} {2018})}\BibitemShut {NoStop}%
\bibitem [{\citenamefont {Schering}\ \emph {et~al.}(2019)\citenamefont
  {Schering}, \citenamefont {Uhrig},\ and\ \citenamefont
  {Smirnov}}]{S_PRRSchering19}%
  \BibitemOpen
  \bibfield  {author} {\bibinfo {author} {\bibfnamefont {P.}~\bibnamefont
  {Schering}}, \bibinfo {author} {\bibfnamefont {G.~S.}\ \bibnamefont
  {Uhrig}},\ and\ \bibinfo {author} {\bibfnamefont {D.~S.}\ \bibnamefont
  {Smirnov}},\ }\bibfield  {title} {\bibinfo {title} {Spin inertia and
  polarization recovery in quantum dots: Role of pumping strength and resonant
  spin amplification},\ }\href
  {https://doi.org/10.1103/PhysRevResearch.1.033189} {\bibfield  {journal}
  {\bibinfo  {journal} {Phys. Rev. Research}\ }\textbf {\bibinfo {volume}
  {1}},\ \bibinfo {pages} {033189} (\bibinfo {year} {2019})}\BibitemShut
  {NoStop}%
\bibitem [{\citenamefont {Leppenen}\ and\ \citenamefont
  {Smirnov}(2022)}]{S_Leppenen2022}%
  \BibitemOpen
  \bibfield  {author} {\bibinfo {author} {\bibfnamefont {N.~V.}\ \bibnamefont
  {Leppenen}}\ and\ \bibinfo {author} {\bibfnamefont {D.~S.}\ \bibnamefont
  {Smirnov}},\ }\bibfield  {title} {\bibinfo {title} {{Optical measurement of
  electron spins in quantum dots: quantum Zeno effects}},\ }\href
  {https://doi.org/10.1039/d2nr01241c} {\bibfield  {journal} {\bibinfo
  {journal} {Nanoscale}\ }\textbf {\bibinfo {volume} {14}},\ \bibinfo {pages}
  {13284} (\bibinfo {year} {2022})}\BibitemShut {NoStop}%
\bibitem [{\citenamefont {Smirnov}\ \emph {et~al.}(2020)\citenamefont
  {Smirnov}, \citenamefont {Zhukov}, \citenamefont {Yakovlev}, \citenamefont
  {Kirstein}, \citenamefont {Bayer},\ and\ \citenamefont
  {Greilich}}]{S_PRBSmirnov20}%
  \BibitemOpen
  \bibfield  {author} {\bibinfo {author} {\bibfnamefont {D.~S.}\ \bibnamefont
  {Smirnov}}, \bibinfo {author} {\bibfnamefont {E.~A.}\ \bibnamefont {Zhukov}},
  \bibinfo {author} {\bibfnamefont {D.~R.}\ \bibnamefont {Yakovlev}}, \bibinfo
  {author} {\bibfnamefont {E.}~\bibnamefont {Kirstein}}, \bibinfo {author}
  {\bibfnamefont {M.}~\bibnamefont {Bayer}},\ and\ \bibinfo {author}
  {\bibfnamefont {A.}~\bibnamefont {Greilich}},\ }\bibfield  {title} {\bibinfo
  {title} {Spin polarization recovery and hanle effect for charge carriers
  interacting with nuclear spins in semiconductors},\ }\href
  {https://doi.org/10.1103/PhysRevB.102.235413} {\bibfield  {journal} {\bibinfo
   {journal} {Phys. Rev. B}\ }\textbf {\bibinfo {volume} {102}},\ \bibinfo
  {pages} {235413} (\bibinfo {year} {2020})}\BibitemShut {NoStop}%
\end{thebibliography}

\let\addcontentsline\oldaddcontentsline
\makeatletter
\renewcommand\tableofcontents{%
    \@starttoc{toc}%
}
\makeatother
\renewcommand{\i}{{\rm i}}


\onecolumngrid
\vspace{\columnsep}
\begin{center}
\newpage
\makeatletter
{\large\bf{Supplemental Material for\\``\@title''}}
\makeatother
\end{center}
\vspace{\columnsep}

This includes the following topics:\\

\twocolumngrid
\hypersetup{linktoc=page}
\tableofcontents
\vspace{\columnsep}

\renewcommand{\thesection}{S\arabic{section}}
\renewcommand{\section}[1]{\oldsec{#1}}
\renewcommand{\thepage}{S\arabic{page}}
\renewcommand{\theequation}{S\arabic{equation}}
\renewcommand{\thefigure}{S\arabic{figure}}
\renewcommand{\bibnumfmt}[1]{[S#1]}
\renewcommand{\citenumfont}[1]{S#1}

\setcounter{page}{1}
\setcounter{section}{0}
\setcounter{equation}{0}
\setcounter{figure}{0}

\section{Supplemental experimental results}

\subsection{Setup and optical properties of samples}

\begin{figure}
  \includegraphics[width=\columnwidth]{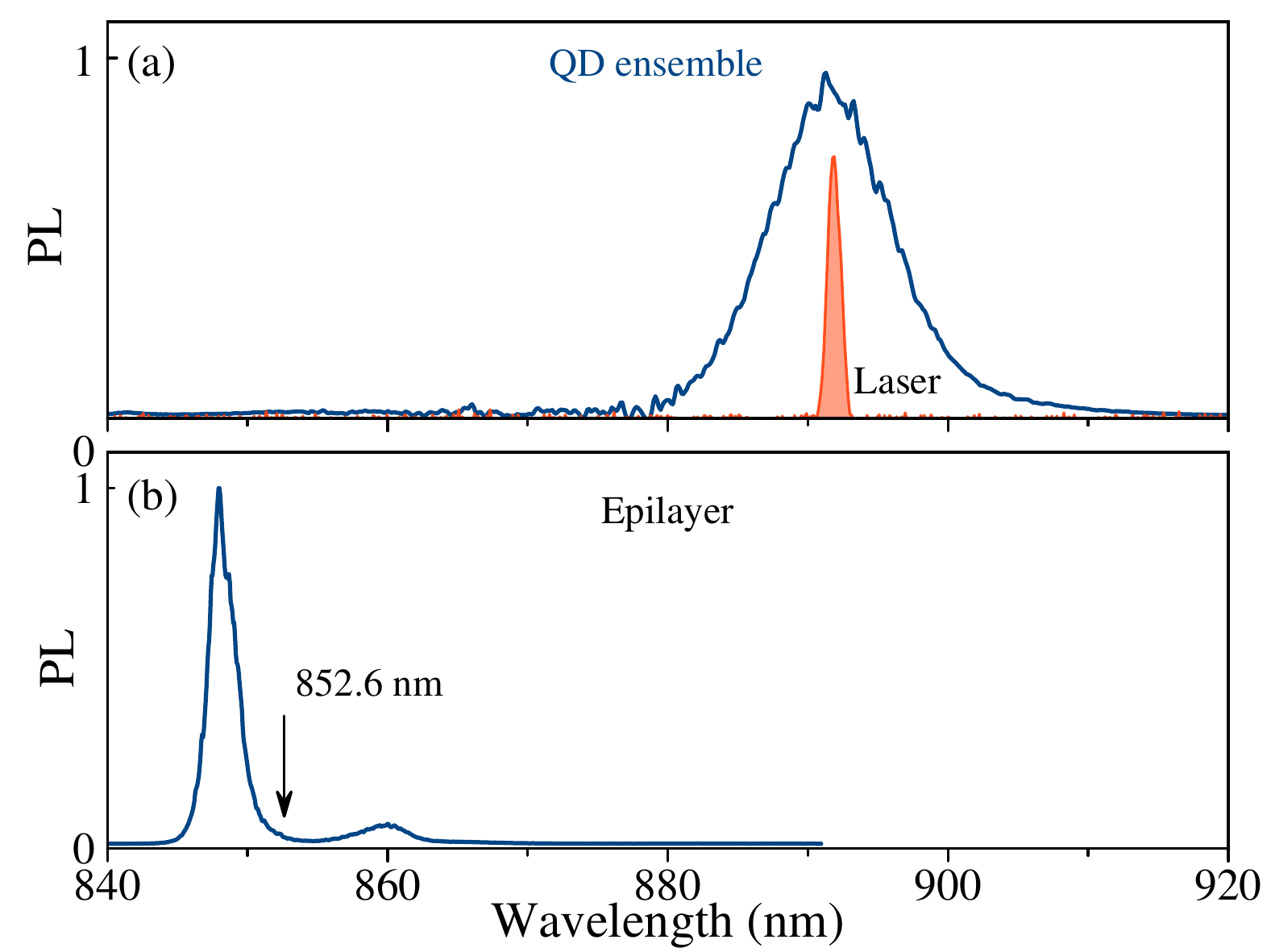}
  \caption{(a) PL of the QD sample (blue) and the laser position (orange). (b) PL of the epilayer and the marked laser position. $T=6\,$K.}
  \label{fig:S1}
\end{figure}

Figure~\ref{fig:S1}(a) and~\ref{fig:S1}(b) demonstrate the photoluminescence (PL) of the QD sample and epilayer used in the main text, respectively. Localized spin-singlet trion complexes are resonantly excited for the QD sample at $\lambda=892\,$nm wavelength at the PL maximum (see Laser in Fig.~\ref{fig:S1}(a)) and for the epilayer sample at $\lambda=852.6\,$nm on the low energy PL flank.

For the QD sample, we use a superconducting magnet, enabling high magnetic fields. For the epilayer sample, we use an electromagnet, which provides low magnetic field strengths with high accuracy. In both cases, the magnets generate a longitudinal magnetic field parallel to the optical $z$ axis (Faraday geometry). The laser pulses with a duration of 2\,ps are emitted with 1\,GHz repetition frequency. We split the laser emission into the pump and probe pulses and delay them relative to each other by a mechanical delay line. A double-modulation scheme is used to reduce the effects of scattered pump and probe light. The pump helicity was modulated between left and right circular polarization by an electro-optical modulator at $f_m^{pu}=10\,$kHz frequency for the QD sample~\cite{S_PRBSchering21} and $f_m^{pu}=84\,$kHz for the epilayer sample. The circularly polarized pump pulses create spin polarization of the resident electrons along the $z$ axis~\cite{S_PRBShabaev03}. The spin polarization accumulates during a series of pulses with the same helicity so that a quasi-steady polarization establishes unless the helicity modulation period is shorter than the electron spin polarization time (spin inertia effect~\cite{S_PRBHeisterkamp15,S_PRBZhukov18,S_PRBSmirnov18,S_PRRSchering19}). The modulation frequencies are thus chosen to be large enough to prevent dynamical nuclear polarization (on the time scale of the nuclear spin polarization time), but sufficiently small to avoid reduction of the spin polarization by the spin inertia effect. The linearly polarized probe beam was modulated in intensity by a photo-elastic modulator in series with a Glan-Thompson prism at a frequency of $f_m^{pr}=100\,$kHz. The signal was detected using a lock-in amplifier at the difference frequency $f_m^{pr}-f_m^{pu}$.

\begin{figure}
  \includegraphics[width=\columnwidth]{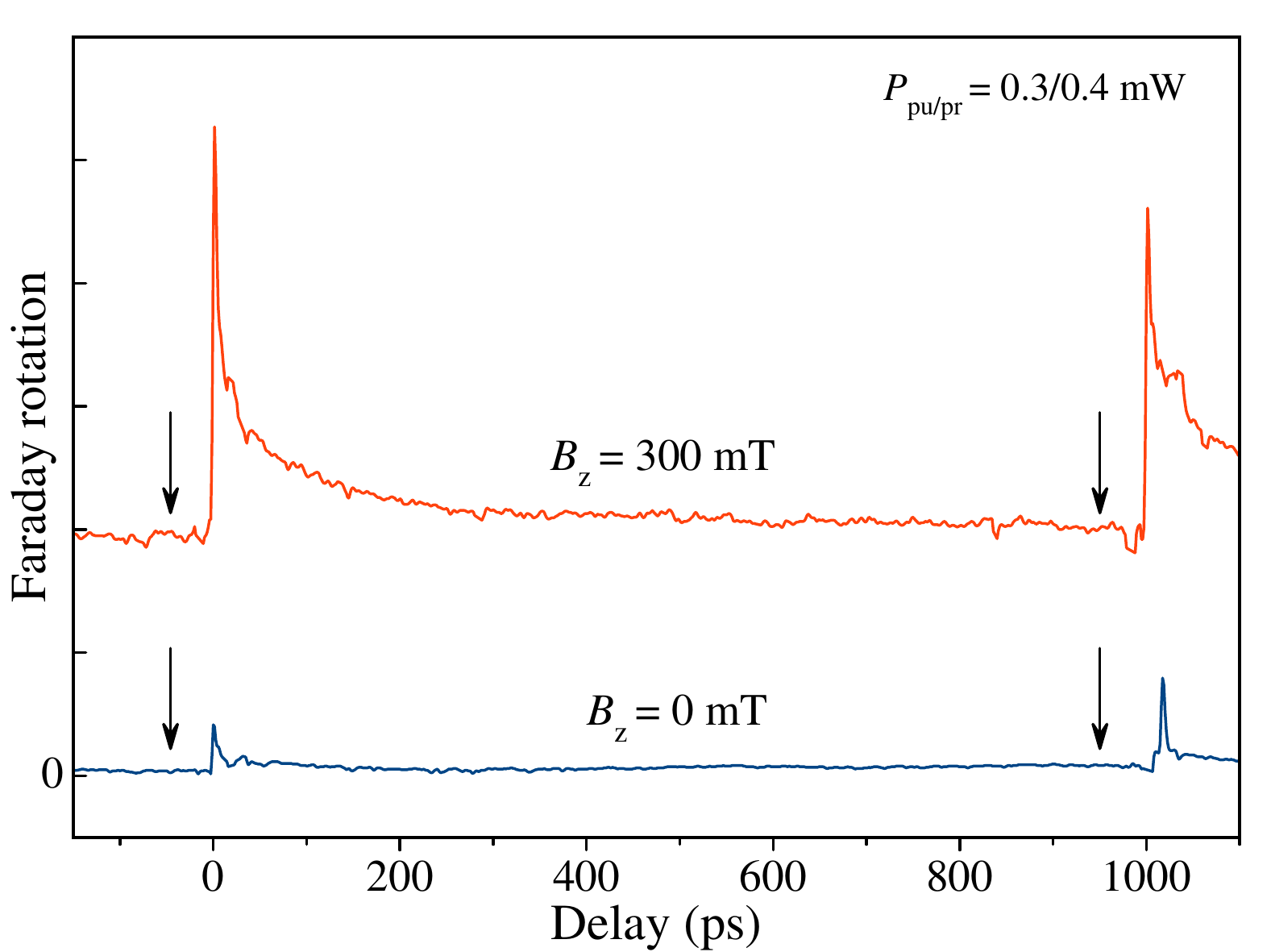}
  \caption{Pump-probe traces for QD sample with $B_z=0$ and $B_z=300\,$mT. Arrows mark the fixed delay of the pump-probe for measurements of PRC. $T=6\,$K.}
  \label{fig:S2}
\end{figure}

Figure~\ref{fig:S2} demonstrates two exemplary time-resolved pump-probe traces for the QD sample measured at two magnetic fields. It demonstrates the increase of the spin polarization seen as an offset of the trace for the higher values of the magnetic field. Further, we fix the delay of pump-probe at $-50$($= +950$\,ps) to measure spin polarization as a function of the longitudinal magnetic field, the PRC. 

\subsection{Additional quantum anti-Zeno effect measurements}

Here we provide results complementary to those in the main text, which demonstrate the anti-Zeno effect for another quantum dot (QD) ensemble named QD2. The sample consists of 20 layers of $n$-doped InGaAs self-assembled quantum dots (QDs) separated by $70$-nm barriers of GaAs. Similar to the QD sample in the main text, a $\delta$-doping of Si at 16\,nm above each layer provides a single electron per QD on average. The rapid thermal annealing at 880\,$^{\circ}\text{C}$ for 30\,s shifts the average emission energy to 1.3662\,eV. The pump-probe measurements are done at this photon energy. The QD density per layer is also about $10^{10}\,$cm$^{-2}$~\cite{S_PRBSchering21}.

\begin{figure}
  \includegraphics[width=\columnwidth]{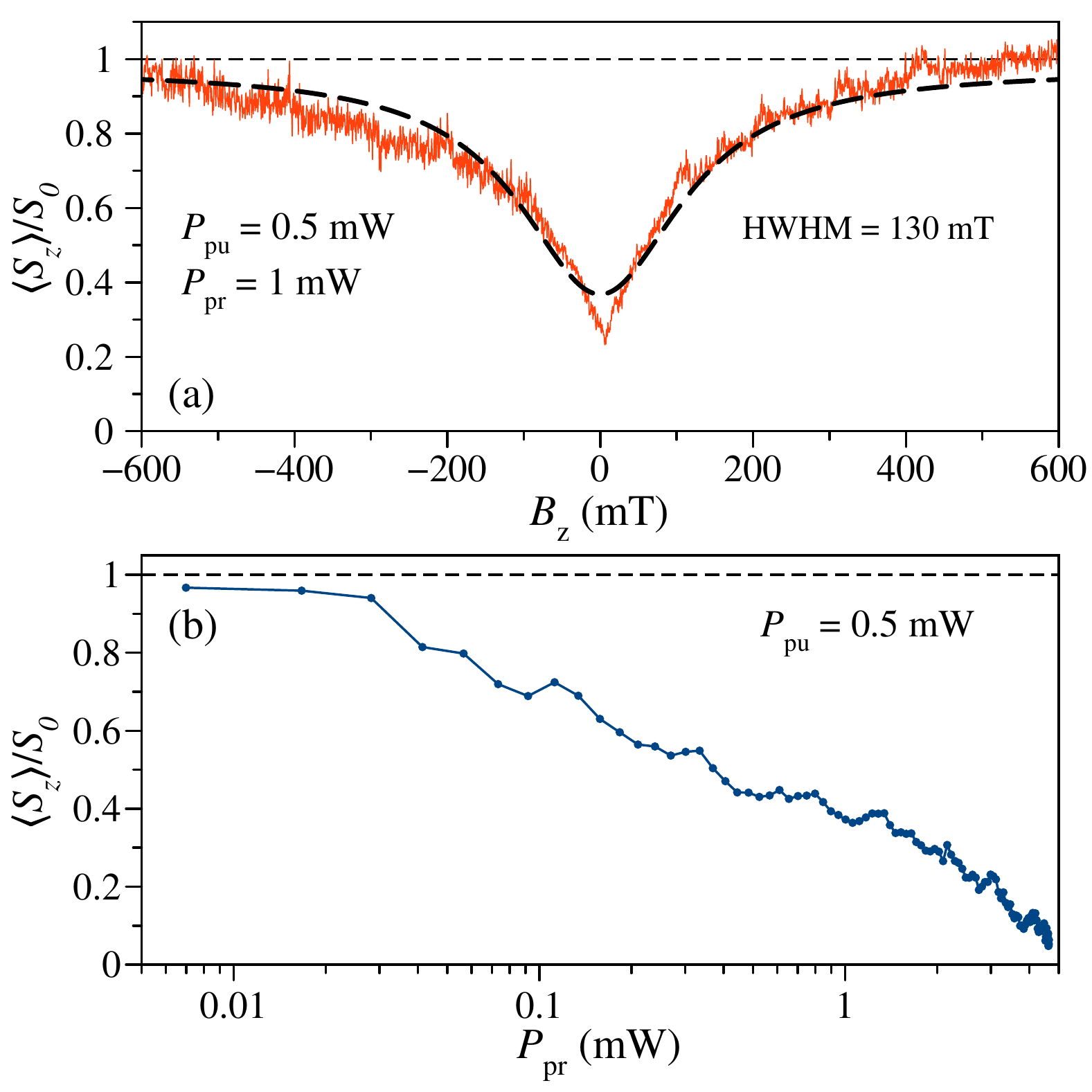}
  \caption{(a) PRC (orange) for the QD2 sample with Lorentzian fit shown by the black dashed curve. (b) Probe power dependence of the zero field spin polarization $\langle S_z \rangle$ normalized by the $S_0$ measured at 1\,T for the pump power $P_\text{pu}=0.5\,$mW. $T=6\,$K.}
  \label{fig:S3}
\end{figure}

Figure~\ref{fig:S3}(a) demonstrates an exemplary PRC measured for the sample QD2 with $P_\text{pu}=0.5\,$mW and $P_\text{pr}=1\,$mW at $f_\text{m}=10\,$kHz. It was reported in Ref.~\cite{S_PRBSchering21}, and its Supplemental Material, that the electron localization volume in this sample is smaller and, accordingly, the hyperfine interaction is stronger. This results in a much wider PRC with a HWHM of 130\,mT. For this sample, the variation of the normalized zero-field spin polarization with probe power [Fig.~\ref{fig:S3}(b)] is much stronger compared to Fig.~\ref{fig:1}(d). This shows the possibility to tune the electron spin relaxation rate by the quantum anti-Zeno effect across a wide range, depending on the sample parameters.

\subsection{Role of pump power in the quantum Zeno effect}

\begin{figure}
  \includegraphics[width=\columnwidth]{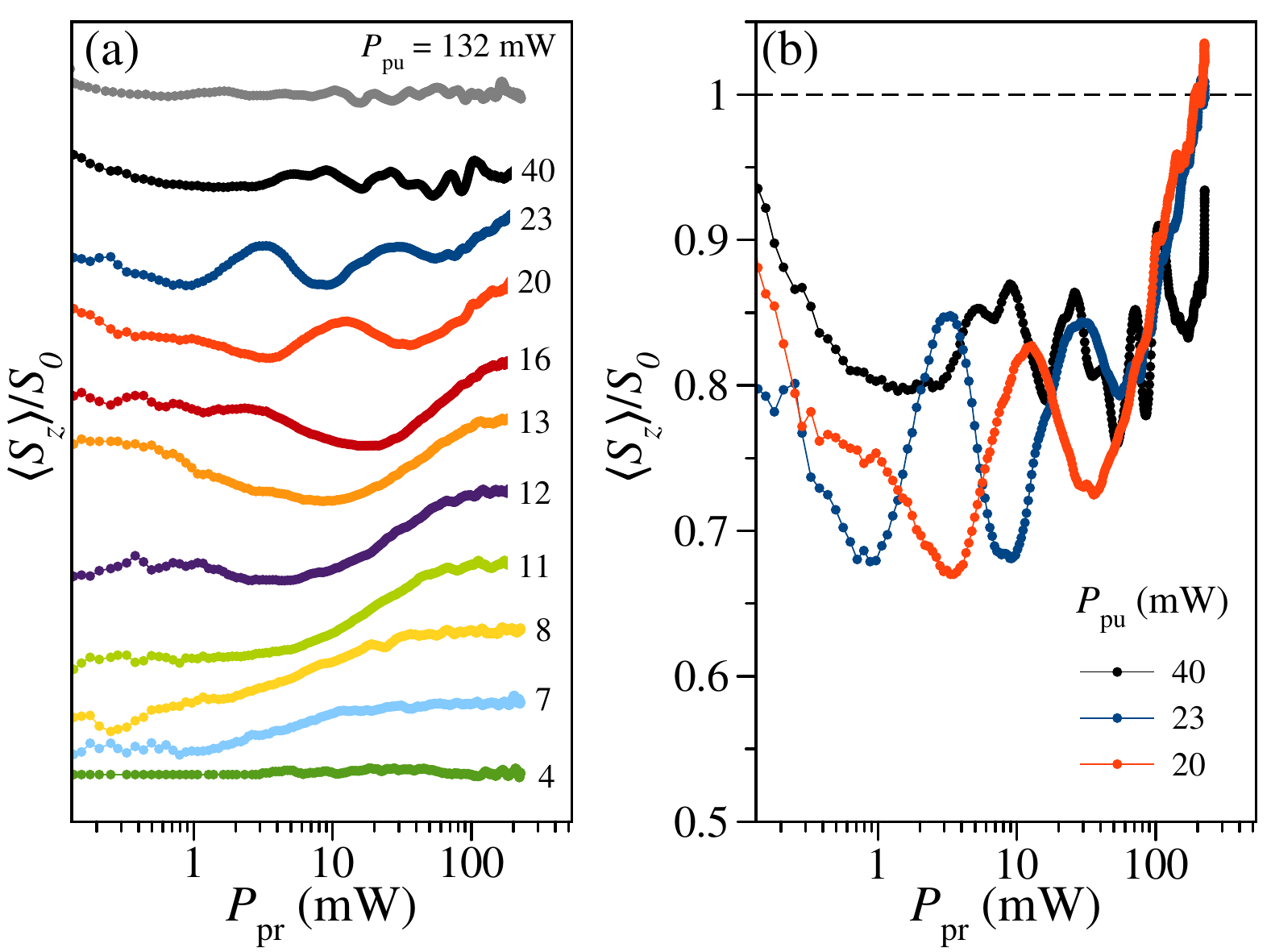}
  \caption{(a) Normalized zero field spin polarization for the epilayer sample for varying probe power at different pump powers. The curves are shifted vertically for better visibility. The pump powers are shown on the right. (b) Exemplary curves for $P_\text{pu} = $20, 23, and 40\,mW, which demonstrate the variation of the oscillation phase.}
  \label{fig:S4}
\end{figure}

Here we present additional measurements for the epilayer sample for variation of the pump power. This experiment was conducted using a laser with a repetition period of $T_\text{R}=13.2\,$ns, which allows for much stronger pulse powers in comparison to the laser with $T_\text{R}=1\,$ns for the same values of the average power measured by a power meter. The values on the axis are normalized to the values of $T_\text{R}=1\,$ns for better comparison (the relation between the probe power and th theoretical parameter $q$ remains approximately the same). Figure~\ref{fig:S4}(a) demonstrates the zero field spin polarization as a function of the probe power at different values of the pump power. For low pump powers, the zero-field spin polarization increases with increasing probe power, in agreement with the previously discussed Zeno effect. At higher values of the pump power, one observes oscillations with varying probe power. These oscillations are related to Rabi rotations, as we show in Sec.~\ref{sec:Rabi}. Figure~\ref{fig:S4}(b) demonstrates several exemplary curves without vertical offset to highlight the variation of the oscillations.

\subsection{Spin inertia measurement}

\begin{figure}[t!]
  \includegraphics[width=\columnwidth]{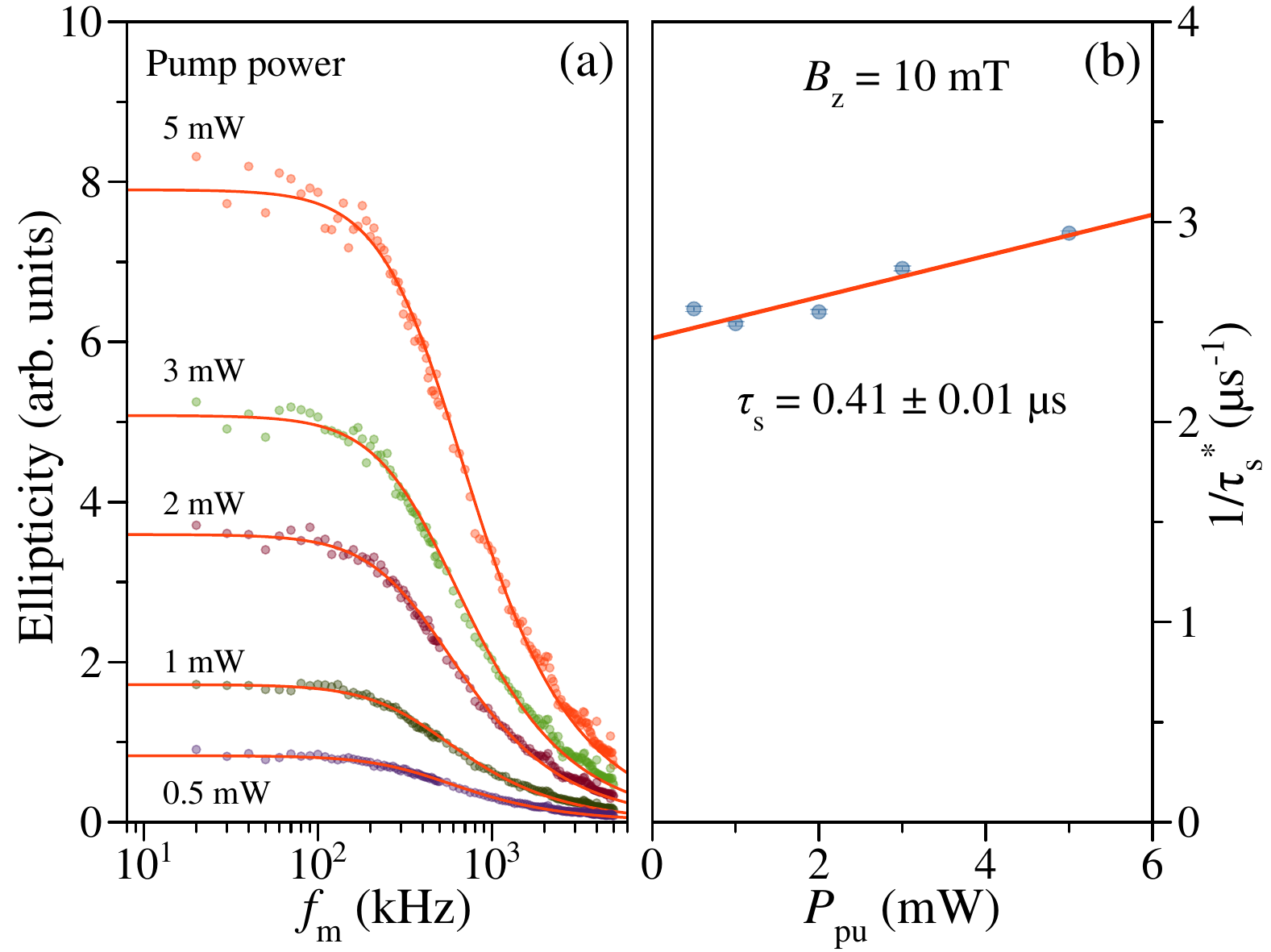}
  \caption{(a) Faraday ellipticity amplitude for the epilayer sample as a function of the pump modulation frequency $f_\text{m}$. It is measured for different pump powers at the magnetic field of $B_\text{z} = 10\,$mT with the pump-probe delay of $-50$\,ps (circles). The lines are fitted by the spin-inertia model. $P_\text{pr} = 0.6\,$mW. (b) Power dependence of the extracted inverse effective spin lifetimes. The linear extrapolation to zero power (orange line) yields $\tau_s = (0.41 \pm 0.01)\,\mu$s.}
  \label{fig:S5}
\end{figure}

Figure~\ref{fig:S5}(a) demonstrates measurements of the spin inertia effect for the epilayer sample~\cite{S_PRBHeisterkamp15}. Here we use the laser with $T_\text{R}=1$\,ns and vary the frequency of the pump helicity modulation $f_\text{m}$ at the fixed longitudinal magnetic field of $B_\text{z}=10\,$mT (much stronger than the HWHM of the PRC) at different pump powers. In this field, the hyperfine interaction is negligible, so the frequency dependence is described by $E(f_\text{m})=E_0/\sqrt{1+(2\pi f_\text{m}\tau_s^*)^2}$, where $\tau_s^*$ is the pump power dependent spin relaxation time unrelated to the electron nuclei interaction. The extracted values of $\tau_s^*$ are plotted in Fig.~\ref{fig:S5}(b) in combination with the extrapolation to zero pump power, which gives the value of the intrinsic longitudinal spin relaxation time~\cite{S_PRBSmirnov18} of $\tau_s=0.41$\,$\mu$s. This time is then used in the modeling of the Zeno effect in the main text.

\section{Theoretical details}

\subsection{Numerical details}

\begin{figure}[t]
  \includegraphics[width=\columnwidth]{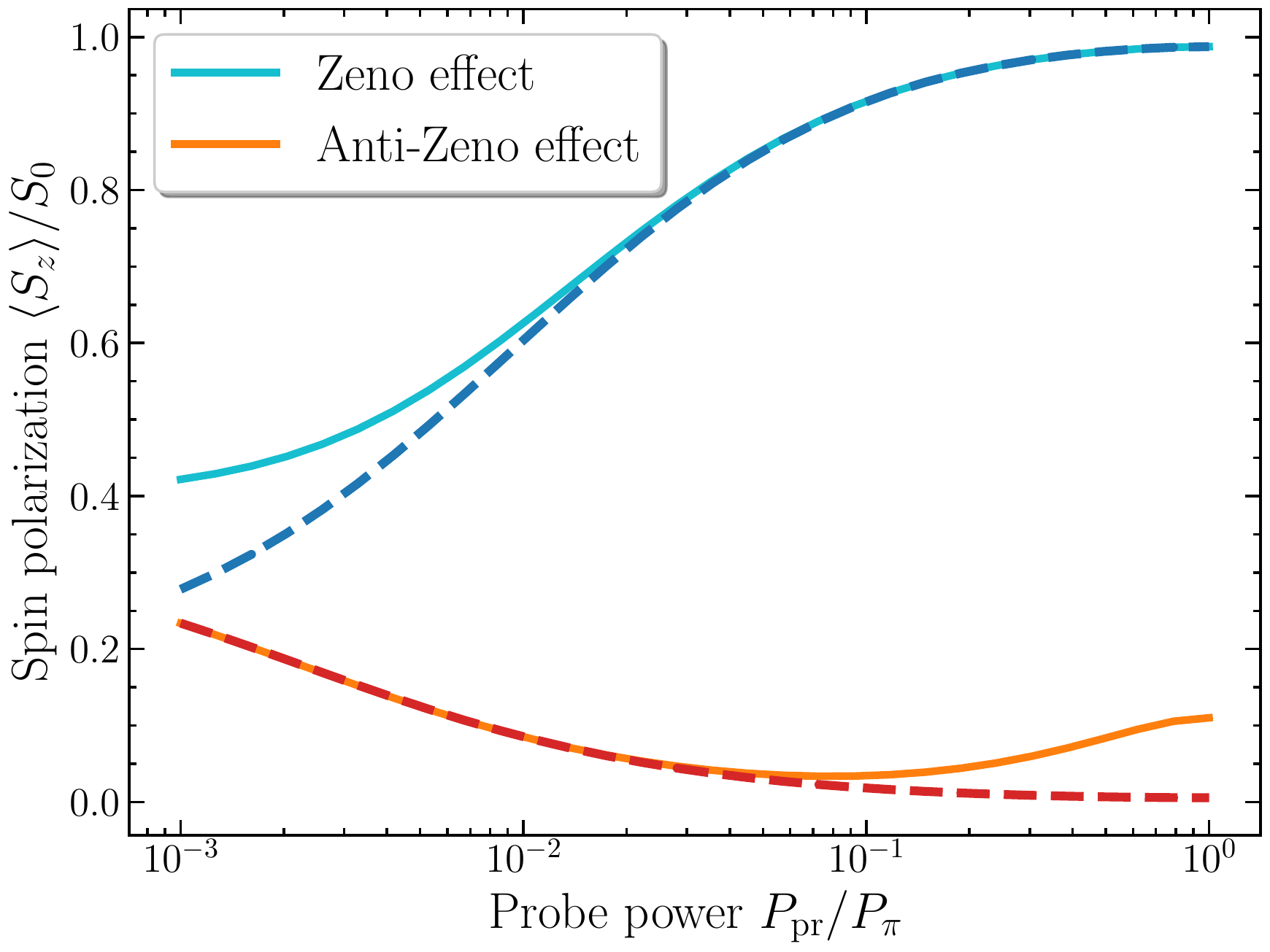}
  \caption{Comparison of the numerical calculations of the spin polarization (cyan and orange curves) with the analytical Eqs.~\eqref{eq:S_z_Zeno} (blue curve) and~\eqref{eq:S_z_AZeno} (red curve). The parameters of the calculation are \addDima{the same as for the Zeno and anti-Zeno effects in the main text}.}
  \label{fig:S6}
\end{figure}

In the steady state, from the solution of Eq.~\eqref{eq:dSt} for the electron spin dynamics, we obtain
\begin{equation}
	\bm S^{-} = \exp(\hat{{\cal M}}T_R)\bm S^{+},
\end{equation}
where the matrix $\hat{\cal M}$ reads
\begin{equation}
\mathcal{M} =	\mqty[-1/\tau_s & -\Omega_{z} &\Omega_{y} \\ \Omega_{z} & -1/\tau_s & -\Omega_x \\ -\Omega_{y} & \Omega_x & -1/\tau_s],
\end{equation}
with $\bm \Omega = \bm \Omega_N +\bm \Omega_L$ being the total electron spin precession frequency. Supplementing this with Eq.~\eqref{eq:cond} which describes the action of the probe and pump pulses, we obtain a system of six linear equations:
  \begin{equation}\label{eq:sup_Mat}
    \mqty[\multicolumn{3}{c}{\multirow{3}{*}{$\exp(\hat{{\cal M}}T_R)$}} &\multicolumn{3}{c}{\multirow{3}{*}{$\mathbf{0}$}} \\ \\ \\
    1 & 0 & 0 & -q^2Q & 0 & 0\\ 0 & 1 & 0 & 0 & -q^2Q & 0\\ 0 & 0 & 1 & 0 & 0 & -1]\mqty[S_x^+ \\ S_y^+ \\ S_z^+ \\ S_x^- \\ S_y^- \\ S_z^-] = \mqty[0 \\ 0 \\ 0 \\ 0 \\ 0\\ g].
  \end{equation}
From its solution, we find the spin $S_{z}^-$, which is measured by the probe pulses, for arbitrary values of $T_R$, $\tau_s$, $\bm\Omega_N$, and $\bm\Omega_L$. This can be done analytically, but the expression is very cumbersome. Next, we average $S_z^-$ over the probability distribution function of the nuclear field
  \begin{equation}
    \mathcal F(\bm\Omega_N)=\frac{\sqrt{2}T_2^*}{\sqrt{\pi}}\e^{-2(\Omega_NT_2^*)^2}.
  \end{equation}
Generally, this can be done only numerically, for example, using the Gauss-Laguerre quadrature scheme, as described in the Supplementary Information of Ref.~\cite{S_Leppenen2022}.

Note that using the expansion of the matrix exponent in the limits $T_R \ll 1/\delta$ and $\tau_s \gg T_R$ up to second order 
\begin{equation}
	\exp(\hat{{\cal M}}T_R) \approx \hat{I}_{3\times 3} + {\cal M}T_R+ \frac{1}{2}({\cal M}T_R)^2
\end{equation}
we rederive Eq.~\eqref{eq:Sz_Zeno} of the main text.

The comparison between the numerical and analytical calculations of the spin polarization for the regime of the quantum anti-Zeno effect is shown in Fig.~\ref{fig:S6} by the orange solid and red dashed curves. \addDima{Here the parameters are the same as in the main text. In the experimentally relevant range of the probe powers, $P_{\rm pr}<0.1P_\pi$ the agreement is very good. At the larger probe powers, the measurement back action is stronger, so the transition between the Zeno and quantum anti-Zeno effects is approached}.

A similar comparison for the regime of the quantum Zeno effect is shown in Fig.~\ref{fig:S6} by the light blue solid and blue dashed curves. The two curves somewhat differ for \addDima{probe power $P_{\rm pr}<0.02P_\pi$.} For this reason, we use the numeric calculation instead of the analytical expression to fit the experiment demonstrating the Zeno effect.

\begin{figure}[t!]
  \includegraphics[width=\columnwidth]{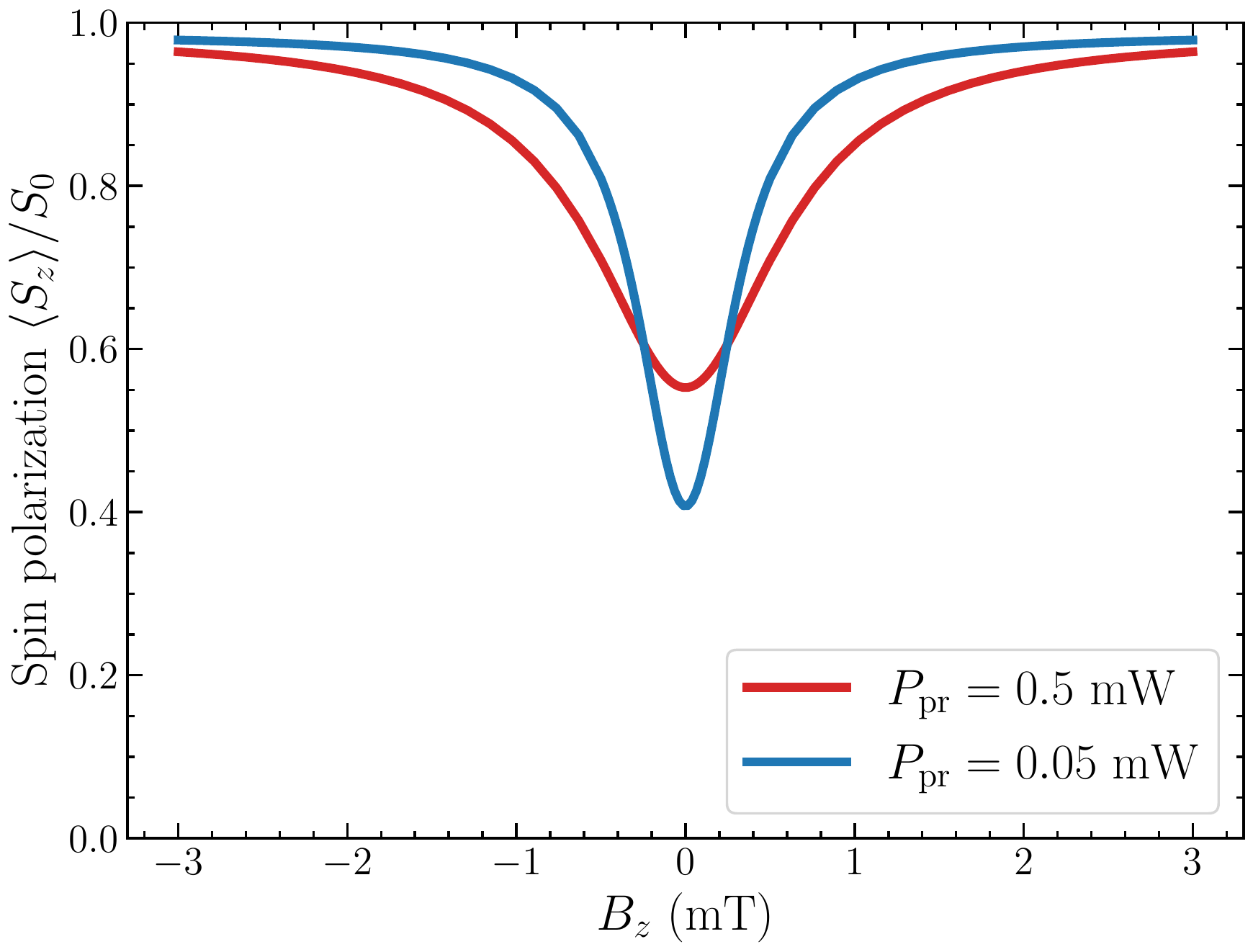}
  \caption{Numerical calculation of PRCs for the parameters $T_2^*=88$\,ns, $\tau_s=0.4\,\mu$s, $P_\pi=80$\,mW, $Q\to 1$, $T_R = 1$\,ns and the two different probe powers given in the legend.
  }
  \label{fig:S7}
\end{figure}

In Fig.~\ref{fig:S7}, we show PRCs calculated numerically for the parameters that fit the quantum-Zeno experiment. They are similar to the experimental data shown in Fig.~\ref{fig:2}(a) but have a slightly smaller HWHM. This may be related to the finite trion lifetime, which is assumed to be zero in our model, or to electron hopping between donors.

\subsection{Role of Rabi oscillations}
\label{sec:Rabi}

We have mainly focused on the case of weak pump and probe pulses assuming the spin polarization to be small and the power of the pulses not to exceed $\pi$. Beyond the former assumption, the relation of the longitudinal spin components before and after the pulses takes the form
\begin{multline}
  S_z^+ = S_z^-+G\left[\frac{1-Q^2}{2}\left(\frac{1}{2}-S_z^-\right)-(1-q^2)S_z^-\right]\\
  +G^2\frac{(1-q^2)(1-Q^2)}{2}S_z^-,
\end{multline}
where $G\in[0;1]$ determines the probability $G/2$ of a trion spin flip during its lifetime~\cite{S_PRBSmirnov20} and the relations for the $S_{x,y}$ components remain the same as in Eq.~\eqref{eq:cond}. In the limit of weak spin pumping, $G\to0$, one recovers Eq.~\eqref{eq:cond} with $g=G(1-Q^2)/4$.

The expressions for the trion excitation probability
\begin{subequations}
  \begin{equation}
    q=\cos\left(\pi/2\sqrt{P_{\text{pr}}/{P_\pi}}\right),
  \end{equation}
  \begin{equation}
    Q=\cos\left(\pi/2\sqrt{P_{\text{pu}}/{P_\pi}}\right),
  \end{equation}
\end{subequations}
can be used for large powers, $P_{\text{pr,pu}}>P_\pi$, while neglecting as well the decoherence during the excitation.

\begin{figure}[t]
  \includegraphics[width=\columnwidth]{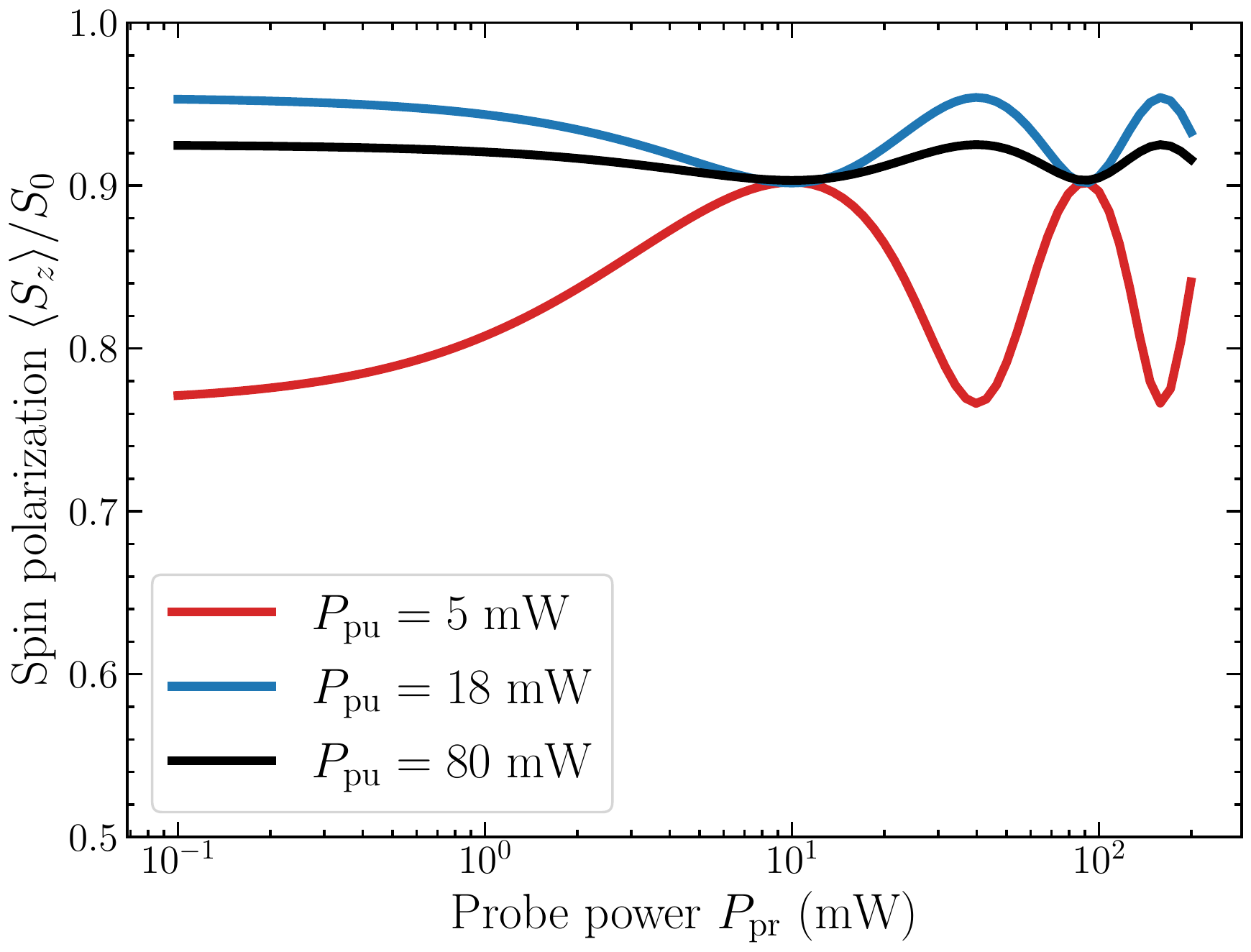}
  \caption{Normalized spin polarization as a function of the probe pulse power calculated numerically for the parameters: $T_2^* = 88$\,ns, $\tau_s =$0.4\,$\mu$s, $T_R = 13.2$\,ns, $P_{\pi} = 10$\,mW, $G=0.1$.
  }
  \label{fig:S8}
\end{figure}

Using this model, we calculate the normalized spin polarization, which is shown in Fig.~\ref{fig:S8}. One can see that this model qualitatively reproduces the experimental data shown in Fig.~\ref{fig:S4}(b). In particular, it explains the oscillations of the spin polarization with probe power as a result of the Rabi oscillations between the electron and trion states for large powers, and the change of the phase of the oscillations because of the additional phase produced by the pump pulse.




%

\end{document}